\documentclass{article}
\usepackage{graphicx} 
\usepackage[letterpaper, left=1in, right=1in, top=0.9in, bottom=0.9in]{geometry}
\usepackage[american]{babel}
\usepackage[normalem]{ulem}
\usepackage{amsmath, amssymb, cases, amsthm}
\usepackage{mdframed,lipsum}
\usepackage{thmtools, thm-restate}
\usepackage{tikz}
\usepackage[shortlabels]{enumitem}
\usepackage{mdframed}
\usepackage{bbm}
\usepackage{bm}
\usepackage{microtype}
\usepackage{xcolor}
\usepackage{makecell}
\usepackage{mathtools}
\usepackage{algorithmic}
\usepackage[procnumbered,ruled,vlined,linesnumbered]{algorithm2e}
\usepackage{float}
\usepackage{varwidth}
\usepackage{svg}
\usepackage{tikz-network}
\usetikzlibrary{decorations.pathreplacing}
\usepackage{subcaption} 
\usepackage{multirow}
\usepackage{natbib} 
\usepackage{tablefootnote}

\RequirePackage{etex} 

\SetKwInput{KwData}{Input}
\SetKwInput{KwResult}{Output}
\SetKwInput{KwGlobalVar}{Global variables}

\usepackage[bookmarks,colorlinks,breaklinks]{hyperref}
\hypersetup{urlcolor=blue, colorlinks=true, citecolor=green!50!black, linkcolor=blue}
\usepackage[capitalize,nosort,nameinlink]{cleveref}
\usepackage{etex}

\DeclareMathOperator*{\argmin}{arg\,min}
\newcommand{\abs}[1]{\vert #1 \vert}
\newcommand{\sset}[1]{\{ #1 \}}

\declaretheorem[numberwithin=section,refname={Theorem,Theorems},Refname={Theorem,Theorems}]{theorem}

\declaretheorem[numberlike=theorem]{lemma}

\declaretheorem[numberlike=theorem]{corollary}
\declaretheorem[numberlike=theorem]{definition}
\declaretheorem[numberlike=theorem]{claim}

\DeclarePairedDelimiter\ceil{\lceil}{\rceil}

\theoremstyle{definition}

\title{Approximate EFX and Exact tEFX Allocations for Indivisible Chores: Improved Algorithms}

\author{Mahyar Afshinmehr\footnote{Department of Computer Engineering, Sharif University of Technology, Iran, mahyarafshinmehr@gmail.com} \and Matin Ansaripour\footnote{Computer and Communications Sciences Department, EPFL, Switzerland, matin.ansaripour@epfl.ch} \and Alireza Danaei\footnote{Department of Computer Engineering, Sharif University of Technology, Iran, alireza.danaei2002@gmail.com} \and Kurt Mehlhorn\footnote{Max Planck Institute for Informatics, SIC, Germany, mehlhorn@mpi-inf.mpg.de}}
\date{September 2024}

\begin{document}

\maketitle

\begin{abstract}
    We explore the fair distribution of a set of $m$ indivisible chores among $n$ agents, where each agent's costs are evaluated using a monotone cost function. Our focus lies on two fairness criteria: envy-freeness up to any item (EFX) and a relaxed notion, namely envy-freeness up to the transfer of any item (tEFX). We demonstrate that a 2-approximate EFX allocation exists and is computable in polynomial time for three agents with subadditive cost functions, improving upon the previous $(2 + \sqrt{6})$ approximation for additive cost functions. This result requires extensive case analysis. ~\cite{ijcai2024p0300} independently claim the same approximation for additive cost functions; however, we provide a counter-example to their algorithm. We expand the number of agents to any number to get the same approximation guarantee with the assumption of partially identical ordering (IDO) for the cost functions. Additionally, we establish that a tEFX allocation is achievable for three agents if one has an additive 2-ratio bounded cost function, while the others may have general monotone cost functions. This is an improvement from the prior requirement of two agents with additive 2-ratio bounded cost functions. This allocation can also be extended to agent groups with identical valuations. Further, we show various analyses of EFX allocations for chores, such as the relaxations for additive $\alpha$-ratio-bounded cost functions.
\end{abstract}

\section{Introduction}

\subsection{Background}

The quest for fairness in the allocation of indivisible items has been a longstanding challenge in the field of algorithmic game theory. The goal of a fair division problem is to allocate a set of goods or chores among agents such that the allocation satisfies certain fairness criteria, such as envy-freeness. These criteria aim to ensure that the allocation is perceived as fair by all agents according to their individual value functions. This area has seen notable advancements, primarily focusing on the concept of envy-freeness, especially under the framework of envy-freeness up to any item (EFX). 

Envy-freeness (EF) is a fundamental concept in the study of fair division, originally proposed by \cite{Foley1967ResourceAA}. An allocation is considered EF if no agent prefers another agent's allocation over his own. EF-allocations do not always exist. Consider a single good and several agents who would all like to have the good. 

Given these limitations, research has pivoted towards the relaxation of EF. A notable relaxation is envy-freeness up to one item (EF1), as established by \cite{Lipton2004OnAF}. Under EF1, any envy between agents must be eliminated by the removal of at most one item from the envied bundle. The existence and polynomial-time computability of EF1 allocations for indivisible items have been confirmed in subsequent studies for goods \cite{Lipton2004OnAF}, chores, and even a mixture of them \cite{bhaskar_et_al:LIPIcs.APPROX/RANDOM.2021.1}. 


\subsection{EFX and Approximate EFX for Goods}

A more robust relaxation of EF is envy-freeness up to any item (EFX), as introduced by \cite{10.1145/3355902}. Under the EFX notion, no agent $i$ would prefer the bundle of another agent $j$ after the removal of any item from $j$'s bundle. Whether an EFX allocation universally exists for goods remains an unresolved and significant open problem in the field.

Extensive studies have focused on special cases and limitations of EFX allocations concerning goods. For instance, considerable progress has been made in scenarios with a limited number of agents. It is known that EFX allocations invariably exist for two agents with general valuations \cite{doi:10.1137/19M124397X} and for three agents with additive valuations \cite{chaudhury2020efx}. However, for four agents, knowledge is limited to the existence of EFX allocations with at most one item unallocated, termed 'charity' \cite{DBLP:journals/corr/abs-2102-10654}.

In cases where the number of agents is not constrained, research has revealed outcomes for particular conditions and restrictive valuation functions. The existence of EFX allocations for goods has been demonstrated when all agents possess identical valuation functions~\cite{doi:10.1137/19M124397X}. This result was further extended to situations where agents are categorized into two groups with identical valuations \cite{MAHARA2023115}. Additionally, EFX allocations have been identified when the number of items exceeds the number of agents by no more than three \cite{mahara2023extension}. 

Regarding these difficulties, research has begun to establish relaxations of EFX allocations. One of the most well-known relaxations is approximate $EFX$. \cite{doi:10.1137/19M124397X} investigated approximate EFX allocations for goods, establishing the existence of a $0.5$-EFX allocation. The approximation factor was later improved by \cite{article} to $1/\phi \approx 0.618$.
\subsection{EFX and Approximate EFX for Chores}
In the setting of chores, little is known about the relaxation of EFX. For two agents, an EFX allocation of chores always exists, by an easy adaption of the argument for goods \cite{doi:10.1137/19M124397X}. Moreover, by \cite{li2022almost}, there exists an EFX allocation when agents share identical ordering in their cost functions. This is also true if all but one agent shares identical ordering in their cost functions and when the number of chores does not exceed twice the number of agents \cite{10.1007/978-3-031-43254-5_15}. 

Transfer EFX (tEFX) is another relaxation we consider in this paper. An allocation $X$ is said to be tEFX for chores if, for any agents $i$ and $j$, the envy of $i$ towards $j$ vanishes upon the transfer of one chore from $X_i$ to $X_j$. \cite{ijcai2023p0277} presents a polynomial-time algorithm for achieving EF1 and Pareto Optimality (PO) allocations with $(n-1)$ surplus in the chores setting. They also provide a polynomial-time algorithm that yields allocations for three agents that are either Proportional or tEFX. For three agents with additive cost functions, \cite{ijcai2022p110} constructs a $(2 + \sqrt{6})$-approximate EFX allocation, and recently \cite{ijcai2024p0300} proved there is no EFX allocation in the case of 3 agents with general monotone cost functions and 6 chores. 


\subsection{Our Contribution}
We study the case where we need to allocate a set of $m$ items to a set of $n$ agents fairly. Our main focus is two relaxations of EFX allocations: approximate EFX and tEFX allocations.

\paragraph{Approximate EFX Allocation.} We give a general framework for approximate EFX allocations on chores for subadditive cost functions. We utilize the work of \cite{bhaskar_et_al:LIPIcs.APPROX/RANDOM.2021.1} that uses a variant of the \textit{Envy Cycle Elimination} technique in the chores setting, namely, \textit{Top Trading Envy Cycle Elimination}. For three agents, we improve the best-known approximation factor of $(2 + \sqrt{6})$ by \cite{ijcai2022p110} to 2 (Theorem~\ref{2-EFX for three agents}), we note that the previous result only holds for additive cost functions where our result extends to subadditive cost functions. It is worth noting that \cite{ijcai2024p0300} independently gave the same framework in order to achieve a 2 approximate EFX allocation for three agents with additive cost functions; however, in Section~\ref{sec: counter example}, we give a counter-example to their algorithm, i.e., give an instance with three agents and 6 items such that no approximation guarantee is achievable with their algorithm. We extend our result to a general number of agents $n$ with subadditive cost functions who agree on the order of the top $n-1$ most costly chores and obtain a 2-EFX allocation (Theorem~\ref{2-EFX for n-partial-IDO}). We next consider approximation guarantees for agents with additive ratio-bounded cost functions where we show the existence of a $(1 + \frac{\alpha - 1}{\lceil m/n \rceil - 1})$-EFX allocation when every agent has an additive $\alpha$-ratio-bounded cost function (Theorem~\ref{alpha ratio theorem}). A direct corollary of this result is the existence of a 2-EFX allocation for agents with additive $\alpha$-ratio-bounded valuation functions when $m \geq \alpha n$. It is worth noting that for a general number of additive agents, only a $(3n^2 - n)$-EFX allocation was known \cite{ijcai2022p110} which was recently improved to a 5-EFX allocation by~\cite{garg2024fairdivisionindivisiblechores}.

\paragraph{tEFX Allocations.} We next focus on tEFX allocations, where we show the existence of such allocations for three agents where at least one of them has an additive 2-ratio-bounded cost function (Theorem~\ref{3 restricted agents tEFX no group}). This improves upon the results of \cite{ijcai2023p0277}, which required at least two of the three agents to have an additive 2-ratio-bounded cost function. We extend our result to a general number of agents that can be divided into three groups based on their cost functions. A group consisting of agents with an identical general monotone cost function, a group with identical additive and 2-ratio-bounded cost function, and a group that is a single agent with a general monotone cost function (Theorem~\ref{3 restricted agents tEFX}). One can easily conclude Theorem~\ref{3 restricted agents tEFX no group} as a corollary of Theorem~\ref{3 restricted agents tEFX} where each group has only one agent. Therefore, we only prove Theorem~\ref{3 restricted agents tEFX} in section \ref{tEFX Existence for a General and 2-ratio-bounded Cost Functions}.

\section{Preliminaries}
For every integer $k$, we denote the set $\{1, 2, \ldots, k\}$ by $[k]$.
We consider a set $[m] = \{1, 2, \dots, m\}$ of indivisible chores and a set $[n] = \{1, 2, \dots, n\}$ of agents. Each agent $i$ has a cost function $C_i: 2^{[m]} \rightarrow \mathbb{R}_{\geq 0}$, which assigns a non-negative cost to each subset of chores. Every cost function $C_i(\cdot)$ is assumed to be a value oracle, e.g., it returns $C_i(S)$ with a computation complexity of $O(1)$ time for every $S\subseteq [m]$. The cost functions are assumed to be monotone, e.g., for each agent $i \in [n]$, $C_i(S \cup \{c\}) \ge C_i(S)$ for all $S \subseteq [m]$ and $c \in [m]$. A cost function $C_i(\cdot)$ is said to be \emph{additive} if $C_i(S) = \sum_{c\in S} C_i(\{c\})$ for all $S \subseteq [m]$, and \emph{subadditive} if $C_i(S \cup T) \leq C_i(S) + C_i(T)$ for all $S, T \subseteq [m]$, where $S\cap T = \emptyset$. An \textit{allocation} of $[m]$ is a vector of sets $X = (X_1,\ldots, X_n)$, where $X_i$ denotes the bundle assigned to agent $i$ ($X_i \subseteq [m]$), for every agent $i \in [n]$, and $X_i \cap X_k = \emptyset$ for every $i \neq k$. For convenience, we use $c$ instead of $\{c\}$ and we write $X_i \setminus c$ and $X_i \cup c$ instead of $X_i \setminus \{c\}$ and $X_i \cup \{c\}$. We call an allocation $X$ a full allocation if all the items in $[m]$ are allocated, i.e., $[m] = \cup_{i = 1}^{n}X_i$, and a partial allocation if at least one item in $[m]$ remains unallocated. We refrain from calling an allocation a full allocation if the context is clear (note that the desired outputs are a full allocation in all the settings mentioned in this paper.)

For each notion of fairness, we say agent $i$ strongly envies agent $j$ if and only if agents $i$ and $j$ violate the fairness criteria, i.e., given an allocation $X = \langle X_1, X_2, \dots, X_n\rangle$, in the context of $\alpha$-EFX allocations, we say that an agent $i$ \emph{strongly envies} an agent $j$ if and only if $C_i(X_i \setminus c) > \alpha C_i(X_j)$ for some $c \in X_i$, and for the context of tEFX allocations, we say that an agent $i$ strongly envies an agent $j$ if and only if $C_i(X_i \setminus c) >  C_i(X_j \cup c)$ for some $c \in X_i$. Thus, an allocation satisfies our fairness criteria if there is no strong envy between any pair of agents. We are now going to introduce some specific definitions and revisit certain concepts that will be crucial in the upcoming sections.

\paragraph{Approximate EFX$(\alpha$-EFX)} For some $\alpha \ge 1$, an allocation is said to be \emph{$\alpha$-Envy-Free up to Any Chore ($\alpha$-EFX)} if for any two agents $i$ and $j$, and for any chore $c$ in the bundle of $i$, it holds that $C_i(X_i \setminus c) \leq \alpha \cdot C_i(X_j)$.

\begin{definition}
\textbf{($\alpha$-EFX allocation)} An allocation $X = (X_1, \ldots, X_n)$ is said to be $\alpha$-Envy-Free up to Any Chore ($\alpha$-EFX) for some $1 \leq \alpha$, if for any two agents $i$ and $j$, and for any chore $c$ in the bundle of $i$, it holds that $C_i(X_i \setminus c) \leq \alpha \cdot C_i(X_j)$, where $C_i(\cdot)$ is the cost function of agent $i$, and $X_i$ and $X_j$ are the sets of chores allocated to agents $i$ and $j$, respectively.    
\end{definition}

\begin{definition}
\textbf{($\alpha$-EFX feasibility)} A bundle $X_i$ in an allocation $X = (X_1, \ldots, X_n)$ is said to be $\alpha$-Envy-Free up to Any Chores (EFX) feasible if for every agent $j \neq i$, and for any chore $c$ in the bundle of $i$, agent $i$ does not envy agent $j$ after the removal of $c$. Formally, it must hold that $C_i(X_i \setminus {c}) \leq \alpha  \cdot C_i(X_j)$ for any chore $c$ in $X_i$.
\end{definition}

\textbf{Envy-freeness up to transferring any chore (tEFX).} An allocation is said to be tEFX if, for any two agents, $i$ and $j$, the envy from agent $i$ towards agent $j$ is eliminated upon the hypothetical transfer of any chore from $i$'s bundle to $j$'s bundle. Formally, this requires that $C_i(X_i \setminus {c}) \leq C_i(X_j \cup {c})$ for any chore $c$ in $X_i$. This definition captures a dynamic aspect of fairness in allocation, emphasizing the ability to resolve envy between agents through the strategic transfer of individual chores. The focus is on the removal of envy via reallocation rather than the comparison of the values of the bundles within a fixed threshold. We also say $i$ tEFX-envies $j$ if there exists $c \in X_i$ such that $C_i(X_i \setminus {c}) > C_i(X_j \cup {c})$.

\begin{definition}
\textbf{(tEFX feasibility)} The allocation $X = \langle X_1, X_2, \dots, X_n\rangle$ is tEFX feasible if for any two agents $i$ and $j$, and for any chore $c$ in the bundle of $i$, $C_i(X_i \setminus {c}) \leq C_i(X_j \cup {c})$.
\end{definition}

\textbf{Cost functions.} Subsequently, we present several key definitions related to cost functions, which play a crucial role in the next sections.
\begin{definition}
We establish the following characteristics for cost functions:
\begin{itemize}
    \item For any $\alpha \ge 1$, a cost function $C(\cdot)$ is \textit{$\alpha$-ratio-bounded} if $$\frac{max_{b \in [m]}C(b)}{min_{b \in [m]}C(b)} \leq \alpha.$$
    \item A set of cost functions $C = \{C_1(\cdot), \ldots, C_n(\cdot)\}$ is \textit{identical ordering (IDO)} if for every two cost functions $C_i(\cdot)$ and $C_j(\cdot)$ in $C$, they share an identical ordering among the cost of chores, i.e., for any $c,c' \in [m]$, $C_i(c) > C_i(c')$ if and only if $C_j(c) >C_j(c')$.
    \item Assume for every cost function $C_i(\cdot)$ in 
 $C$, $T(C_i) = \langle c_1, c_2, \dots, c_m \rangle$ which for every $t < s$, $C_i(c_t) > C_i(c_s)$. A set of cost functions $C$ is \textit{$k$-partial identical ordering ($k$-partial-IDO)} if for every two cost functions $C_i(\cdot)$ and $C_j(\cdot)$ in $C$, they share an identical ordering among the cost of $k$ chores with the highest costs, i.e., for any $1 \le t \le \min(k, m)$, $T(C_i)_t = T(C_j)_t$. 
\end{itemize}
\end{definition}

\textbf{Non-degenerate instances.} We extend the concept of non-degenerate instances, as introduced by \cite{chaudhury2020efx}. 
\cite{DBLP:journals/corr/abs-2205-07638} demonstrated that to prove the existence of EFX allocations for goods where agents possess general monotone valuations, it is sufficient to establish the existence of EFX allocations solely for non-degenerate instances. We adapt their approach to show a similar corollary in the chores setting.

We define an instance of cost functions $C = \{C_1(\cdot), \dots, C_n(\cdot)\}$ as non-degenerate if, for any agent $i \in [n]$, the cost function $C_i$ satisfies $C_i(S) \neq C_i(T)$ for all distinct sets of chores $S$ and $T$. This definition ensures that each agent perceives a unique cost for every possible bundle of chores, thereby eliminating the possibility of cost ties between different sets. Throughout our study, we proceed under the assumption that we are dealing with non-degenerate instances. This assumption implies that all agents assign a positive marginal cost to every chore. Detailed proof of the sufficiency of focusing on non-degenerate instances is provided in Appendix \ref{appen_non_degenerate}.

\begin{algorithm}[tb]
\caption{Top trading envy cycle elimination algorithm} \label{TTECE}
\KwIn{$M :$ set of chores, $N :$ set of agents, $C :$ set of subadditive cost functions}
\KwOut{Allocation $X$}

$X \gets \{\emptyset, \ldots, \emptyset\}$

\While{$\exists$ unallocated item $c$}{
\While{$\not\exists$ a sink in the top trading envy graph for partial allocation $X$}{
Find an envy cycle $i_1 \rightarrow i_2 \rightarrow \ldots \rightarrow i_t \rightarrow i_1$ in the top trading envy graph with respect to $X$.

\For{$1 \leq k \leq t$}{
$X_{i_k} \gets X_{i_{(k+1) \mod t}}$}

}
Choose sink $s$ in the top trading envy graph with respect to $X$.

$X_s \gets X_s \cup \{c\}$
}

\Return{$X$}

\end{algorithm}

\textbf{Top-Trading Envy-Graph.} One of the main tools we use to derive EFX approximations is the concept of the top trading envy graph, which resembles a sub-graph of the well-known envy graph as introduced in ~\cite{Lipton2004OnAF}. We note that the concept of the top trading envy graph is the same as in \cite{bhaskar_et_al:LIPIcs.APPROX/RANDOM.2021.1}; however, its application to achieve approximate EFX allocations is our key focus in our framework.

\begin{definition}[Top Trading Envy Graph]
    For a partial allocation $X = (X_1, \ldots, X_n)$ in the setting with $n$ agents and $m$ chores, the top-trading envy-graph is as follows: consider a graph with $n$ vertices, one for each agent, we add a directed edge from agent $i$ to agent $j$ if and only if we have $C_i(X_i) > C_i(X_j)$ and $C_i(X_j) = \min_{k \in [n]}(C_i(X_k))$.
\end{definition}

We wish to reach an allocation with a node with no output edges in the top trading envy graph, so that then by adding chores to such agent we can eventually reach an approximate EFX allocation. Therefore, we use the top trading envy cycle elimination procedure to remove top trading envy cycles in the top trading envy graph as in \cite{bhaskar_et_al:LIPIcs.APPROX/RANDOM.2021.1} (Algorithm \ref{TTECE}). This procedure is similar to the envy cycle elimination procedure on the envy graph in the goods setting.


We utilize the top-trading envy-graph to apply the cycle elimination technique as shown in detail in section \ref{2-EFX Existence for three agents and n-partial-IDO Cost Functions}. The key lemma involves a partial $\alpha$-EFX allocation. Consider if there is some $\beta > 0$ such that for any unallocated item $b$ and any agent $i$, the following holds: there are at least $n-1$ agents $j$ for whom $C_i(b) \leq \beta \cdot C_i(X_j)$. Under these conditions, we can employ the cycle-elimination procedure on the top trading envy graph. This method allows us to allocate the remaining items. Consequently, we return a full $\max(\alpha, \beta + 1)$-EFX allocation.

\section{2-EFX existence for three agents and $(n-1)$-partial-IDO Cost Functions}\label{2-EFX Existence for three agents and n-partial-IDO Cost Functions}

In this section, we show the existence of a 2-EFX allocation for three agents with subadditive cost functions and for any number $n$ of agents with subadditive and $(n-1)$-partial-IDO cost functions; additionally, we give a counter-example for the algorithm in~\cite{ijcai2024p0300} which claims to return a 2-EFX allocation for three agents with additive cost functions. Our results rely on the following lemmas, which show how to extend a partial approximate EFX allocation using cycle elimination. 

\begin{lemma} \label{envy cycle elimination in top-trading envy graoh}
 Let $X$ be an \(\alpha\)-EFX allocation. After the cycle elimination step (the inner while loop of Algorithm \ref{TTECE}) on the top trading envy graph, the resulting allocation remains $\alpha$-EFX. Additionally, the procedure runs in polynomial time.
\end{lemma}

\begin{proof}
    It is clear that we only need to show after we remove one cycle via the top trading cycle elimination procedure, the resulting allocation remains $\alpha$-EFX. Based on the definition of the top trading envy graph, if the allocation of an agent changes, it will receive the minimum cost bundle with respect to that agent's cost function among all the allocated bundles; therefore, it will not envy any other agent. Also, the set of bundles does not change, and the allocated bundles of agents not in the top trading envy cycle will not change. Thus, we can clearly see the resulting allocation remains $\alpha$-EFX.

    In each iteration, we reduce the number of edges in the top trading envy graph. Since the outdegree of any agent in the top trading envy graph is at most one, our algorithm will terminate after, at most, $n$ rounds of cycle elimination. In each round, we spend a polynomial time swapping the bundles and checking whether a cycle remains. Therefore, the envy cycle elimination in the top trading envy graph runs in polynomial time.
\end{proof}

In the following, we adapt a similar statement as \cite{articlesedighin}, which is for goods. Compared to the case of the goods, we must search for an EFX allocation with stronger assumptions to use the following lemma to obtain an approximate EFX allocation for chores. 

\begin{lemma} \label{max(alpha, (beta + 1))-EFX chores}
Consider the setting with a set of agents $[n]$ having subadditive cost functions. Let $X = (X_1, X_2, ..., X_n)$ be a partial \(\alpha\)-EFX allocation and $P$ the set of unallocated chores. There exists a $\max(\alpha, \beta + 1)$-EFX full allocation like $X'$ if for a $\beta > 0$ and every agent \(i\), we have a set of at least $n-1$ agents $R_i(X)$, such that for any $j \in R_i(X)$ and unallocated chore \(b \in P\), \(C_i(b) \leq \beta \cdot C_i(X_j)\). If so, this allocation can be found in polynomial time, given access to a value oracle for the subadditive valuation functions.
\end{lemma}

\begin{proof}
While \(P \neq \emptyset \), we let \(b \in P\) be an arbitrary unallocated chore. In each iteration of our algorithm, we update \(X\) such that \(b\) will be allocated, and the resulting new allocation remains $\max(\alpha, (\beta + 1))$-EFX. Note that at the beginning of our algorithm, the allocation \(X\) is $\max(\alpha, (\beta + 1))$-EFX. We first remove all cycles in the top trading envy graph corresponding to the partial allocation \(X\). Since the resulting graph has no cycles, we can choose a sink \(i_0\), i.e., an agent who does not envy any other agent, and add \(b\) to this agent's bundle. This procedure is represented via Algorithm \ref{TTECE}. We will now prove that the resulting allocation remains $\max(\alpha, \beta + 1)$-EFX.

By Lemma~\ref{envy cycle elimination in top-trading envy graoh}, after the cycle elimination step, the resulting allocation is still $\max(\alpha, \beta + 1)$-EFX. Since \(i_0\) is a sink, we have \(C_{i_0}(X_{i_0}) \leq C_{i_0}(X_{j})\) for every agent \(j\). Moreover, since in our algorithm, we only add chores to the allocated bundles, if we have \(C_{i_0}(b) > \beta C_{i_0}(X_{j})\) for some $j \neq i_0$, we will have \(C_{i_0}(b) > \beta C_{i_0}(X_{j}) \ge \beta C_{i_0}(X_{i_0})\) which contradicts our assumption that  \(C_{i_0}(b) \leq \beta C_{i_0}(X_{j})\) for at least $n-1$ agents $j$, i.e., agents $j \in R_i(X)$. So we can assume \(C_{i_0}(b) \leq \beta C_{i_0}(X_{j})\) for any $j \neq i_0$. By the subadditivity of our valuation functions, for any $j \neq i_0$, we have 

\begin{align*}
    C_{i_0}(X_{i_0} \cup b) &\leq C_{i_0}(X_{i_0}) + C_{i_0}(b)\\ 
    &\leq C_{i_0}(X_{j}) + \beta C_{i_0}(X_{j})  = (1 + \beta) C_{i_0}(X_{j}).
\end{align*}

As the other bundles remain unchanged and only the bundle of \(X_{i_0}\) has increased, we can conclude that our new allocation is $\max(\alpha, \beta + 1)$-EFX. Since in each iteration, the size of \(P\) will decrease, the algorithm will terminate, and we return \(X'\), the allocation obtained after allocating the last unallocated item, which is $\max(\alpha, \beta + 1)$-EFX.

We argue that our algorithm runs in polynomial time. By Lemma~\ref{envy cycle elimination in top-trading envy graoh}, the cycle elimination step runs in polynomial time, and in each round, the number of unallocated chores decreases by 1. Therefore, we only need to run the cycle elimination step for at most $m$ rounds. Finding the sink also only takes polynomial time. Therefore, our algorithm will terminate after polynomial time, which concludes the proof of Lemma~\ref{max(alpha, (beta + 1))-EFX chores}. 
\end{proof}


We now show that a 2-EFX allocation exists for $n$ agents with subadditive $(n-1)$-partial-IDO cost functions. 

\begin{theorem} \label{2-EFX for n-partial-IDO}
A $2$-EFX allocation always exists for $n$ agents with subadditive $(n-1)$-partial-IDO cost functions.
\end{theorem}

\begin{proof}
     We first arbitrarily allocate the $n-1$ most costly chores among the agents such that each agent receives at most one chore and one agent remains empty. Call this partial allocation $X = (X_1, ..., X_n)$. Clearly, $X$ is EFX since each agent receives at most one item. Also, since the cost functions are $(n-1)$-partial-IDO, for any unallocated chore $b$ and any agent $i$, there exist $n-1$ agents $j$ such that $C_i(b) \le C_i(X_j)$, namely the agents that have exactly one chore. Therefore, by inserting partial allocation $X$ and parameters $\alpha = \beta = 1$ to Lemma~\ref{max(alpha, (beta + 1))-EFX chores}, we obtain a full allocation $X$ that is 2-EFX. Clearly, the first partial allocation can be found in poly-time since we only need to sort the chores based on the cost function of an arbitrary agent. The algorithm in Lemma~\ref{max(alpha, (beta + 1))-EFX chores} also runs in polynomial time. Therefore, we can find such allocation in polynomial time.
\end{proof}

\subsection{2-EFX Allocation for Three Agents with Subadditive Cost Functions}

 We will now show a 2-EFX allocation always exists for three agents with subadditive cost functions.

 \begin{theorem} \label{2-EFX for three agents}
A 2-EFX allocation always exists among three agents, each with a subadditive cost function.
\end{theorem}

In order to prove Theorem~\ref{2-EFX for three agents}, we aim to construct a partial allocation $X$ satisfying the following properties: 
\begin{enumerate}
    \item $X$ is a 2-EFX allocation,
    \item for every unallocated chore $b$ and agent $i \in [3]$, we have $C_i(b) \le C_i(X_j)$ for at least two $j \in [3]$.
\end{enumerate}
Assuming we can find such a partial allocation $X$ satisfying properties 1 and 2, by inserting the partial allocation $X$ and parameters $\alpha = 2$ and $ \beta = 1$ in Lemma~\ref{max(alpha, (beta + 1))-EFX chores}, we can construct a full $2$-EFX allocation in polynomial time given access to a value oracle for the subadditive cost functions of the three agents.

Note that it has been shown in \cite{ijcai2024p0300} that an EFX allocation exists when $m \le n + 2$ for agents with general monotone cost functions; since $n = 3$, when $m \leq 5$, we can find an EFX allocation in polynomial time given access to a value oracle for the subadditive cost functions, thus, we assume $m \geq 6$. Our approach is based on allocating the two most costly chores with respect to each agent among two distinct agents such that the resulting allocation is 2-EFX. Clearly, if we can achieve such an allocation, we have satisfied both properties. We split the problem into cases based on the intersections of the three agents' most costly chores. It is worth noting that for some cases, notably when two agents have IDO cost functions, we directly construct a full 2-EFX allocation.

\subsection{Constructing Partial Allocation Satisfying Properties (1) and (2) or a Full 2-EFX Allocation}
 
 We show how to construct a partial allocation $X$ in polynomial time satisfying properties (1) and (2) or a full allocation satisfying property (1).
 For any $i \in [3]$ and $t \in [m]$, we define $c^{t}_i$ as the $t$'th most costly chore with respect to $C_i$. Also, as mentioned, we can assume $m \ge 6$. We will use this assumption many times in our arguments without directly mentioning it. To construct the partial allocation satisfying (1) and (2), we consider the most costly chores with respect to the three agents and distinguish cases according to their overlap, e.g., if two agents share the most costly chore. We also make use of the following fact. 

 \begin{lemma}\label{costly are split} Let $i \in [3]$. If the two most costly chores, according to $C_i$, are assigned to distinct agents, (2) holds for $i$. \end{lemma} 
 \begin{proof} Clearly for any of those two agents, say $j$, and any unallocated item $b$, we have $C_i(X_j) \geq C_i(c^2_i) \geq C_i(b)$. The first inequality holds since at least one of $c^1_i$ and $c^2_i$ is allocated to agent $j$, and the second inequality holds since $b$ is not among the two most costly chores with respect to $C_i$.  \end{proof}


Now, we state our different cases and give a complete analysis for them.

\paragraph{Case A, $\abs{\sset{c^1_1,c^1_2,c^1_3}}= 1$:} We divide the problem further based on the second most costly chore.

\paragraph{Case A1, $\abs{\sset{c^2_1,c^2_2,c^2_3}} = 1$:} We set the bundles of agents 1, 2, and 3 to be $\{c^{3}_{1}, c^{3}_{2}, c^{3}_{3}\}$, $\{c^{2}_{1}\}$, and $\{c^{1}_{1}\}$ respectively. Property (2) holds according to Lemma~\ref{costly are split}. Agents 2 and 3 receive singletons, so they do not strongly envy any of the agents. Additionally,

\begin{align*}
    \max_{c \in X_1} C_1(X_1 \setminus c) = &\max_{c, c' \in X_1} C_1(\{c, c'\}) \leq \max_{c, c' \in X_1} C_1(c) + C_1(c')\\ &\leq 2C_1(c^2_1) = 2C_1(X_2) \leq 2C_1(c^1_1) = 2C_1(X_3).
\end{align*}

Where the first inequality comes from the subadditivity of the cost functions, and the second and third inequality comes from the fact that $c^1_1$ and $c^2_1$ are, respectively, the highest and second highest costly chores with respect to $C_1$. Therefore, agent 1 will not strongly envy agents 2 and 3; thus, the allocation is 2-EFX. So (1) holds.

\paragraph{Case A2, $\abs{\sset{c^{2}_1, c^{2}_2, c^{2}_3}} = 2$:} W.l.o.g assume $c^{2}_1 =  c^{2}_2 \neq  c^{2}_3$. We allocate $c^1_1$ to agent 1, $c^2_1$ to agent 3, and let agent 2 have the bundle $\{c^{3}_{1}, c^{3}_{2}, c^{2}_{3}\}$. Property (2) holds by Lemma~\ref{costly are split}. 1 and 3 have only a single item and hence do not strongly envy any other agent. Additionally,

\begin{align*}
    \max_{c \in X_2} C_2(X_2 \setminus c) = &\max_{c, c' \in X_2} C_2(\{c, c'\}) \leq \max_{c, c' \in X_2} C_2(c) + C_2(c')\\ &\leq 2C_2(c^2_2) = 2C_2(X_3) \leq 2C_2(c^1_2) = 2C_2(X_1).
\end{align*}

Where the first inequality comes from the subadditivity of the cost functions, and the second and third inequality comes from the fact that $c^1_2$ and $c^2_2$ are, respectively, the highest and second highest costly chores with respect to $C_2$. Therefore, agent 2 will not strongly envy agents 1 and 3; thus, the allocation is 2-EFX. So (1) holds.

Agent 2 owns at most three chores, each of which is less costly than the chore owned by either agent 1 or 3. So, the allocation is 2-EFX.

\paragraph{Case A3, $\abs{\sset{c^{2}_1, c^{2}_2, c^{2}_3}} = 3$:}  We allocate $c^1_1$ to agent 3, $c^2_2$ to agent 1 and $c^2_1$ to agent 2. We then let agents 1 and 2 choose the most costly remaining chore with respect to $C_3$; it does not matter whether 1 or 2 chooses first. Property (2) holds by Lemma~\ref{costly are split}.  Agent 3 has only a single item and, therefore, does not strongly envy any other agent. Agent 1 and 2 own at most two chores. Since agents 3 and 2 have the two most costly chores with respect to $C_1$ and agents 1 and 3 have the two most costly chores with respect to $C_2$, agents 1 and 2 will not strongly envy any other agent, and therefore, property (1) holds.

\paragraph{Case B, there exists agents $i, j$ such that $\sset{c^1_i, c^2_i} = \sset{c^1_j, c^2_j}$:} W.l.o.g assume $\{c^1_1, c^2_1\} = \{c^1_2, c^2_2\}$. We again divide the problem into sub-cases based on the other most costly chores.

\paragraph{Case B1, B and $\sset{c^{1}_3, c^2_3} \cap \sset{c^{1}_1, c^{2}_1} \not= \emptyset$:} We allocate $c^1_1$ to agent 3 and $c^2_1$ to agent 2 . We then let agent 1 choose the most costly remaining chore with respect to $C_3$. Clearly, the two most costly chores with respect to any agent are allocated to two distinct bundles, so property (2) holds. Since all bundles are singletons, the allocation is EFX. Therefore, property (1) holds.

\paragraph{Case B2, B and $\sset{c^{1}_3, c^2_3} \cap \sset{c^{1}_1, c^{2}_1} = \emptyset$:} We split further.

\paragraph{Case B21, B2 and $c^{3}_1 \ne c^3_2$:} We allocate $c^1_1$ and $c^2_1$ to agent 3, $c^3_1$ to agent 2, and $c^3_2$ to agent 1. Since the two most costly chores with respect to $C_3$ are not equal to $c^1_1$ and $c^2_1$, if they are not been allocated, we can allocate these two chores between agents 1 and 2 such that each of them has exactly one of $c^1_3$ and $c^2_3$. Clearly, the two most costly chores with respect to any agent are allocated to two distinct bundles, so property (2) holds. Since we distributed the two most costly chores with respect to $C_3$ between agents 1 and 2, and since agent 3 has 2 chores, agent 3 will not strongly envy any other agent. One can observe that agents 1 and 2 also do not strongly envy any other agent. Therefore, property (1) holds.

\paragraph{Case B22, B2 and $c^{3}_1 = c^3_2$: } We split further. 

\paragraph{Case B221: B22 and $c^3_1 \in \sset{c_3^1,c_3^2}$: } We allocate $c^1_1$ and $c^2_1$ to agent 3, $c^1_3$ to agent 2, and $c^2_3$ to agent 1. Agents 1 and 2 receive a single item. Therefore, they will not strongly envy any other agent. Also, for any $c \in X_3$, we have $C_3(X_3 \setminus c) \le \min(C_3(X_1), C_3(X_2))$ and our allocation is 2-EFX since agents 1 and 2 have the two most costly chores of agent 3. So, property (1) holds. Since one of $X_1$ or $X_2$ contains $c^3_1$ and $c^3_1 = c^3_2$, property (2) holds.

\paragraph{Case B222: B22 and $c^3_1 \not\in \sset{c_3^1,c_3^2}$: } We split further.

\paragraph{Case B2221, B222 and $c^1_1=c^1_2$ and $c^2_1=c^2_2$.} 
Let $b^1_1$ and $b^2_1$ be, respectively, the maximum and minimum valued chore with respect to $C_1$ among $c^1_3$ and $c^2_3$. Define $b^1_2$ and $b^2_2$ similarly with respect to $C_2$. Allocate $c^1_1$ to agent 3, $b^2_1$ and $c^2_1$ to agent 1, and $b^1_1$ to agent 2 (we can do so since, by the case definition, these are distinct chores). Note that there are $m' = m - 4 \geq 2$ unallocated items at this point. Let $M'$ denote the set of unallocated items. We aim to find a subset of unallocated chores like $D \subseteq M'$ such that for any item $d \in D$ we have  

\begin{align}\label{eq: property D}
    C_1(b^1_1 \cup D) \geq C_1(c^2_1) > C_1(b^1_1 \cup (D \setminus d)).
\end{align}

Assume

\begin{align}\label{eq: assumption 1}
    C_1(b^1_1 \cup M') \geq C_1(c^2_1),
\end{align}
    in this case, we can find a set $D \subseteq M'$ satisfying Equation~\ref{eq: property D} in polynomial time given access to a value oracle for the subadditive cost functions. 

 \begin{lemma}\label{lemma: finding subset D}
     Assuming $m \geq 6$, the conditions of case B2221, and equation~\ref{eq: assumption 1} holds, one can find a non-empty subset $D$ of $M'$ such that for any item $d \in D$ equation~\ref{eq: property D} holds. Additionally, this subset can be found in polynomial time, given access to a value oracle for the subadditive cost functions.
 \end{lemma}

 \begin{proof}
     We first set $D$ to $M'$. Note that at this point we have $C_1(b^1_1 \cup D) > C_1(c^2_1)$ by equation~\ref{eq: assumption 1}. If there exists some item $d' \in D$ such that $C_1(b^1_1 \cup (D \setminus d')) > C_1(c^2_1)$, we remove $d'$ from $D$, i.e., we set $D$ to be $D \setminus d'$. Note that at that point, we still have $C_1(b^1_1 \cup D)> C_1(c^2_1)$. We continue this procedure until no such item $d' \in D$ exists such that $C_1(b^1_1 \cup (D \setminus d')) > C_1(c^2_1)$. We will eventually reach such state since for $D = \emptyset$ we have that 
     
     \begin{align*}
         C_1(b^1_1 \cup D) = C_1(b^1_1) \leq C_1(c^2_1) 
     \end{align*}
    where the inequality comes from the fact that $b^1_1 \notin \{c^1_1, c^2_1\}$, i.e.,  from the case conditions. Therefore, we will eventually reach a subset $D \subseteq M'$ satisfying equation~\ref{eq: property D} for any item $d \in D$. 

    Given access to a value oracle for the cost functions, we can check for the existence of such item $d'$ in $O(|D|)$ time at each step of the algorithm. Additionally, we have at most $m'$ steps in our algorithm to reach the desired set $D$. Therefore, the runtime is at most 
    
    \begin{align*}
         O(\sum_{i = 1}^{m'} i) = O(m'^2) = O(m^2).
    \end{align*}

    Thus, we can find such subset $D \subseteq M'$ satisfying equation~\ref{eq: property D} in polynomial time given access to a value oracle for the subadditive cost functions. 
 \end{proof}

 By Lemma~\ref{lemma: finding subset D}, we can find a non-empty subset $D \subseteq M'$ satisfying equation~\ref{eq: property D} for every item $d \in D$. We allocate this set to agent 2. Property (2) clearly holds. Agent 3 is singleton and hence does not strongly envy any other agent. Agent 1 does not strongly envy agent 2 since

 \begin{align*}
     \max_{c \in X_1} C_1(X_1 \setminus c) = C_1(c^2_1) \leq C_1(b^1_1 \cup D) = C_1(X_2),
 \end{align*}
 where the inequality comes from the fact that equation~\ref{eq: property D} holds. Similarly, agent 1 also does not envy agent 3 since
 
 \begin{align*}
     \max_{c \in X_1} C_1(X_1 \setminus c) = C_1(c^2_1) \leq C_1(c^1_1) = C_1(X_3).
 \end{align*}
 where the inequality again coms from the fact that equation~\ref{eq: property D} holds. If agent 2 does not envy agent 1, we have 

\begin{align*}
    C_2(X_2) &\leq C_2(X_1) = C_2(b^2_1 \cup c^2_1) \\
    &\leq C_2(b^2_1) + C_2(c^2_1) \leq 2C_2(c^1_1) = 2C_2(X_3),
\end{align*}
    therefore, agent 2 will not strongly envy agent 1 or 3 and both properties will hold. 

 If agent 2 envies agent 1, i.e., 
 
 \begin{align}\label{eq: other case 1}
     C_2(b^1_1 \cup D) > C_2(b^2_1 \cup c^2_1),
 \end{align}
 we allocate $\{b^1_1\} \cup D$ to agent 1 and $\{b^2_1, c^2_1\}$ to agent 2. Agent 3 is still singleton and, hence, does not strongly envy any other agent. Clearly, agent 2 does not envy agent 1 since equation~\ref{eq: other case 1} holds, and does not strongly envy agent 3 since

 \begin{align*}
     \max_{c \in X_2}C_2(X_2 \setminus c) = C_2(c^2_1) \leq C_2(c^1_1) = C_2(X_3).
 \end{align*}
 
  We now argue that agent 1 does not strongly envy the other agents. The item with minimum marginal value in $X_1$ with respect to $C_1$ is either $b^1_1$ or some $d \in D$. If this item is some $d \in D$ we have 
 \begin{align*}
     \max_{c \in X_1}C_1(X_1 \setminus c) &= C_1(X_1 \setminus d) = C_1(b^1_1 \cup (D \setminus d))  \\
     &\leq C_1(c^2_1) \leq C_1(X_2) = C_1(b^2_1 \cup c^2_1) \\
     &\leq C_1(b^2_1) + C_1(c^2_1) \leq 2C_1(c^1_1) = 2C_1(X_3).
 \end{align*}
 
Where the first inequality comes from equation~\ref{eq: property D}, the second and third inequality comes from the subadditivity of the cost functions, and the fourth inequality comes from the fact that $c^1_1$ is the most costly chore with respect to $C_1$. Thus, in this case, agent 1 does not strongly envy any of agents 2 or 3, and property (1) holds. 

If $b^1_1$ is the item with minimum marginal value in $X_1$ respect to $C_1$, we have

\begin{align}\label{eq: property 2}
    \max_{c \in X_1}C_1(X_1 \setminus c) = C_1(X_1 \setminus b_1^1) = C_1(D).
\end{align}

We will now show in this case $C_1(D) \leq 2C_1(c^2_1)$.
\begin{lemma}\label{lemma: helper 1}
    Assuming $m \geq 6$, the conditions of case B2221, equation~\ref{eq: assumption 1}, and~\ref{eq: other case 1} holds, we will have 

    \begin{align*}
        C_1(D) \leq 2C_1(c^2_1).
    \end{align*}
\end{lemma}

\begin{proof}
Assume the contrary that 

\begin{align}\label{eq: contrary assumption}
        C_1(D) > 2C_1(c^2_1).
    \end{align}

    By equation~\ref{eq: property D} for any item $d \in D$ we have
    
    \begin{align}\label{eq: helper 1}
    C_1(c^2_1) \ge C_1(b^1_1 \cup (D \setminus d)),
    \end{align}
 adding $C_1(d)$ to both sides of equation~\ref{eq: helper 1} and from the subadditivity of the cost functions, we will have

\begin{align}\label{eq: helper 2}
    C_1(c^2_1) + C_1(d) \geq C_1(b^1_1 \cup (D \setminus d)) + C_1(d) \geq C_1(b^1_1 \cup D),
\end{align}
from equations~\ref{eq: contrary assumption} and~\ref{eq: helper 2}, we have that

\begin{align*}
    C_1(D) + C_1(c^2_1) + C_1(d) \geq C_1(b^1_1 \cup D) + 2C_1(c^2_1),
\end{align*}
which simplifies to 

\begin{align}\label{eq: helper 3}
    C_1(D) + C_1(d) \geq C_1(b^1_1 \cup D) + C_1(c^2_1).
\end{align}

However, since the cost functions are subadditive, we have 

\begin{align}\label{eq: helper 4}
    C_1(D) < C_1(b^1_1 \cup D),
\end{align}
and since $d \notin \{c^1_1, c^2_1\}$, we have

\begin{align}\label{eq: helper 5}
    C_1(d) < C_1(c^2_1).
\end{align}

By combining equations~\ref{eq: helper 4} and~\ref{eq: helper 5}, we have 

\begin{align}\label{eq: helper 6}
    C_1(D) + C_1(d) < C_1(b^1_1 \cup D) + C_1(c^2_1).
\end{align}

Clearly equation~\ref{eq: helper 3} and~\ref{eq: helper 6} contradict each other and we have shown that we have $C_1(D) \leq 2C_1(c^2_1)$ which concludes the proof of this lemma.
\end{proof}

We now have that

\begin{align*}
    \max_{c \in X_1}C_1(X_1 \setminus c) = C_1(D) \leq 2C_1(c^2_1) \leq 2C_1(\{b^2_1, c^2_1\}) = 2C_1(X_2),
\end{align*}
where the first equality comes from equation~\ref{eq: property 2}, the first inequality comes from Lemma~\ref{lemma: helper 1}, and the second inequality comes from the subadditivity of the cost functions. Therefore, agent 1 does not strongly envy agent 2. Similarly, we have 

\begin{align*}
    \max_{c \in X_1}C_1(X_1 \setminus c) = C_1(D) \leq 2C_1(c^2_1) \leq 2C_1(c^1_1) = 2C_1(X_3),
\end{align*}
where the first equality comes from equation~\ref{eq: property 2}, the first inequality comes from Lemma~\ref{lemma: helper 1}, and the second inequality comes from the fact that $c^1_1$ is the most costly chore with respect to $C_1$. Therefore, agent 1 does not strongly envy agent 3. We conclude that in this case property (1) will hold.

We now have to consider the case where equation~\ref{eq: assumption 1} does not hold, i.e., we have

\begin{align}\label{eq: assumption 2}
    C_1(b^1_1 \cup M') < C_1(X_1)
\end{align}



 We allocate all the remaining items to agent 2. Then agent 1 envies agent 2. Agent 3 owns a singleton and does not strongly envy any agent. 

If agent 2 envies agent 1, we swap the bundles of agents 1 and 2 and so remove any envy between these agents. Since 

\begin{align*}
    C_1(X_1) &\leq C_1(c^2_1 \cup b^2_1) \leq C_1(c^2_1) + C_1(b^2_1) \\
    &\leq 2C_1(c^2_1) \leq 2C_1(c^1_1) = 2C_2(X_3),
\end{align*}
where the first inequality comes from equation~\ref{eq: assumption 2}, the second inequality comes from the subadditivity of the cost functions, and the third and fourth inequality come from the fact that $c^1_1$ and $c^2_1$ are respectively the first and second most costly chore with respect to $C_1$. Similarly, 

\begin{align*}
    C_2(X_2) &= C_2(c^2_1 \cup b^2_1) \leq C_2(c^2_1) + C_2(b^2_1) \\
    &\leq 2C_2(c^2_1) \leq 2C_2(c^1_1) = 2C_2(X_3),
\end{align*}
where the first inequality comes from the subadditivity of the cost functions, and the third and fourth inequality come from the fact that $c^1_1$ and $c^2_1$ are respectively the first and second most costly chore with respect to $C_1$. We now have reached a full 2-EFX allocation. 

If agent 2 does not envy agent 1, it also does not strongly envy agent 3 since 

\begin{align*}
    2C_2(X_3) &= 2C_2(c^1_1) \geq C_2(c^2_1) + C_2(b^2_1) \\
    &\geq C_2(c^2_1 \cup b^2_1) = C_2(X_1) \geq C_2(X_2),
\end{align*} 
where the first inequality comes from the fact that $c^1_1$ is the most costly chore with respect to $C_2$ and the second inequality comes from the subadditivity of the cost functions.

If agent 1 does not strongly envy any other agent, we have a full 2-EFX allocation. So let us assume agent 1 envies some other agent. It is clear that agent 1 does not strongly envy agent 3 since

\begin{align*}
    \max_{c \in X_1}C_1(X_1 \setminus c) = C_1(c^2_1) \leq C_1(c^1_1) = C_1(X_3),
\end{align*}
where the inequality comes from the fact that $c^1_1$ is the most costly chore with respect to $C_1$. Assume agent 1 strongly envies agent 2, i.e., since $\max_{c \in X_1} C_1(X_1 \setminus c) = C_1(c^2_1)$, we have 

\begin{align}\label{eq: helper 7}
    C_1(c^2_1) > 2C_1(b^1_1 \cup M').
\end{align}







Consider the new full allocation where agent 1 receives $\{b^1_1 \cup b^2_1 \cup M'\}$, agent 2 receives $c^2_1$, and agent 3 receives $c^1_1$. Agents 2 and 3 own singletons and do not strongly envy any other agents. We have 

\begin{align*}
    C_1(b^1_1 \cup b^2_1 \cup M') \le 2C_1(c^2_1),
\end{align*}
since if we assume the contrary, i.e., $C_1(b^1_1 \cup b^2_1 \cup M') > 2C_1(c^2_1)$, and combining with the inequality $2C_1(c^2_1) > 4C_1(b^1_1 \cup M')$ , i.e., equation~\ref{eq: helper 7}, we have  

\begin{align*}
    C_1(b^1_1 \cup b^2_1 \cup M') > 4C_2(b^1_1 \cup M'),
\end{align*}
and by the subadditivity of the cost functions we have 

\begin{align*}
    C_1(b^2_1) + C_1(b^1_1 \cup M') \geq C_1(b^1_1 \cup b^2_1 \cup M') > 4C_2(b^1_1 \cup M'),
\end{align*}
simplifying to

\begin{align*}
    C_1(b^2_1) > 3C_2(b^1_1 \cup M'),
\end{align*}
 which is clearly a contradiction since $c^3_1 \in M'$ and $b^2_1 \notin \{c^1_1, c^2_1\}$, i.e., 
 \begin{align*}
     C_1(b^2_1) \leq C_1(c^3_1) \leq C_1(b^1_1 \cup M') \leq 3C_1(b^1_1 \cup M').
 \end{align*}

Therefore, we return this full 2-EFX allocation. 

\paragraph{Case B2222, B222 and $c^1_1=c^2_2$ and $c^2_1=c^1_2$.} 
Let $b^1_1$ and $b^2_1$ be, respectively, the maximum and minimum valued chore with respect to $C_1$ among $c^1_3$ and $c^2_3$. Define $b^1_2$ and $b^2_2$ similarly with respect to $C_2$. Allocate $c^2_2$ to agent 3, $b^2_1$ and $c^1_2$ to agent 1, and $b^1_1$ to agent 2 (We can do so since by the case definition, these are distinct chores). Note that by Lemma~\ref{costly are split}, property (2) holds. Since agents 2 and 3 have a single item allocated to them, they will not strongly envy any other agent. Therefore, if agent 1 does not strongly envy agents 2 and 3, we are done. Note that since

\begin{align*}
    C_1(X_2) = C_1(b^1_1) < C_1(c^2_2) = C_1(X_3),
\end{align*}
if agent 1 does not strongly envy agent 2, it will not strongly envy agent 3. Hence, we assume the contrary, i.e. 

\begin{align}\label{eq: helper 8}
    C_1(c_2^1) > 2C_1(b_1^1).
\end{align} 

Let $M'$ be the set of unallocated items. Note that $m' = m - 4 \geq 2$ unallocated items are left. We now follow a similar procedure as case B2221, i.e., we aim to find a subset of unallocated chores like $D \subseteq M'$ such that for any item $d \in D$ we have

\begin{align}\label{eq: property D2}
    C_1(b^1_1 \cup D) \geq C_1(c^1_2) > C_1(b^1_1 \cup (D \setminus d)).
\end{align}

Assume

\begin{align}\label{eq: assumption 3}
    C_1(b^1_1 \cup M') \geq C_1(c^1_2),
\end{align}
    in this case, we can find a set $D \subseteq M'$ satisfying equation~\ref{eq: property D2} in polynomial time given access to a value oracle for the subadditive cost functions. 

 \begin{lemma}\label{lemma: finding subset D2}
     Assuming $m \geq 6$, the conditions of case B2222, and equation~\ref{eq: assumption 3} holds, one can find a non-empty subset $D$ of $M'$ such that for any item $d \in D$ equation~\ref{eq: property D2} holds. Additionally, this subset can be found in polynomial time, given access to a value oracle for the subadditive cost functions.
 \end{lemma}

 \begin{proof}
     We first set $D$ to $M'$. Note that at this point we have $C_1(b^1_1 \cup D) \ge C_1(c^1_2)$ by equation~\ref{eq: assumption 3}. If there exists some item $d' \in D$ such that $C_1(b^1_1 \cup (D \setminus d')) \ge C_1(c^1_2)$, we remove $d'$ from $D$, i.e., we set $D$ to be $D \setminus d'$. Note that at that point, we still have $C_1(b^1_1 \cup D) \ge C_1(c^1_2)$. We continue this procedure until no such item $d' \in D$ exists such that $C_1(b^1_1 \cup (D \setminus d')) \ge C_1(c^1_2)$. We will eventually reach such state since for $D = \emptyset$ we have that 
     
     \begin{align*}
         C_1(b^1_1 \cup D) = C_1(b^1_1) < C_1(c^1_2) 
     \end{align*}
    where the inequality comes from the fact that $b^1_1 \notin \{c^1_1, c^2_1\}$, i.e., from the case conditions. Therefore, we will eventually reach a subset $D \subseteq M'$ satisfying equation~\ref{eq: property D2} for any item $d \in D$. 

    Given access to a value oracle for the cost functions, we can check for the existence of such item $d'$ in $O(|D|)$ time at each step of the algorithm. Additionally, we have at most $m'$ steps in our algorithm to reach the desired set $D$. Therefore, the runtime is at most 
    
    \begin{align*}
         O(\sum_{i = 1}^{m'} i) = O(m'^2) = O(m^2).
    \end{align*}

    Thus, we can find such subset $D \subseteq M'$ satisfying equation~\ref{eq: property D2} in polynomial time given access to a value oracle for the subadditive cost functions. 
 \end{proof}

By Lemma~\ref{lemma: finding subset D2}, we can find a non-empty subset $D \subseteq M'$ satisfying equation~\ref{eq: property D2} for every item $d \in D$. We allocate this set to agent 2. Property (2) clearly holds. Agent 3 is singleton and hence does not strongly envy any other agent. Agent 1 does not strongly envy agent 2 since

 \begin{align*}
     \max_{c \in X_1} C_1(X_1 \setminus c) = C_1(c^1_2) \leq C_1(b^1_1 \cup D) = C_1(X_2),
 \end{align*}
 where the inequality comes from the fact that equation~\ref{eq: property D2} holds. Similarly, agent 1 also does not strongly envy agent 3 since
 
 \begin{align*}
     \max_{c \in X_1} C_1(X_1 \setminus c) = C_1(c^1_2) \leq C_1(c^2_2) = C_1(X_3).
 \end{align*}
 where the inequality again comes from the fact that equation~\ref{eq: property D2} holds. If agent 2 does not envy agents 1 and 3, property (1) will hold. Therefore, assume the contrary. We will make separate case arguments for when agent 2 envies agent 1 or not.

 If agent 2 envies agent 1, i.e., 

\begin{align}\label{eq: helper 9}
    C_2(b^1_1 \cup D) > C_2(b^2_1 \cup c^1_2),
\end{align}
we consider the new allocation below

\begin{align*}
    X = \langle\{b^1_1\} \cup D, \{c^2_2, b^2_1\}, \{c^1_2\}\rangle.
\end{align*}

Property (2) holds from Lemma~\ref{costly are split}. Agent 3 is singleton and does not strongly envy any other agent. Agent 2 does not strongly envy agent 1 since

\begin{align*}
    \max_{c \in X_2} C_2(X_2 \setminus c) &= C_2(c^2_2) \leq C_2(c^1_2) \leq C_2(c^1_2 \cup b^2_1) \\
    &< C_2(b^1_1 \cup D) = C_2(X_1),
\end{align*}
Where the first inequality comes from the fact that $c^1_2$ and $c^2_2$ are respectively the first and second most costly chore with respect to $C_2$, the second inequality comes from the monotonicity of the cost functions, and the third inequality comes from equation~\ref{eq: helper 9}. Similarly, agent 2 will not strongly envy agent 3 since

\begin{align*}
    \max_{c \in X_2} C_2(X_2 \setminus c) = C_2(c^2_2) \leq C_2(c^1_2) = C_2(X_3),
\end{align*}
Where again, the inequality comes from the fact that $c^1_2$ and $c^2_2$ are respectively the first and second most costly chore with respect to $C_2$.

We now show that agent 1 does not strongly envy agents 2 and 3. First note that the item with minimum marginal value in $X_1$ with respect to $C_1$ is either $b^1_1$ or some $d \in D$. If this item is some $d \in D$ we have 

\begin{align*}
    \max_{c \in X_1} C_1(X_1 \setminus c) &= C_1(b^1_1 \cup (D \setminus d)) < C_1(c^1_2) \\
    &= C_1(X_3) < C_1(c^2_2 \cup b^2_1) = C_1(X_2),
\end{align*}
where the first inequality comes from equation~\ref{eq: property D2}, the second inequality comes form the fact that $c^2_2$ and $c^1_2$ are respectively the highest and second highest valued item with respect to $C_1$. Therefore, agent 1 will not envy any of agents 2 or 3. Now assume the item with minimum marginal value in $X_1$ with respect to $C_1$ is $b^1_1$. Similar to what we do in case B2221, we will show that $C_1(D) \leq 2C_1(c^1_2)$.

\begin{lemma}\label{lemma: helper 2}
    Assuming $m \geq 6$, the conditions of case B2222 and equation~\ref{eq: assumption 3} holds, we will have 

    \begin{align*}
        C_1(D) \leq 2C_1(c^1_2).
    \end{align*}
\end{lemma}

\begin{proof}
Assume the contrary that 

\begin{align}\label{eq: contrary assumption 2}
        C_1(D) > 2C_1(c^1_2).
    \end{align}

    By equation~\ref{eq: property D2} for any item $d \in D$ we have
    
    \begin{align}\label{eq: helper 10}
    C_1(c^1_2) \ge C_1(b^1_1 \cup (D \setminus d)),
    \end{align}
 adding $C_1(d)$ to both sides of equation~\ref{eq: helper 10} and from the subadditivity of the cost functions, we will have

\begin{align}\label{eq: helper 11}
    C_1(c^1_2) + C_1(d) \geq C_1(b^1_1 \cup (D \setminus d)) + C_1(d) \geq C_1(b^1_1 \cup D),
\end{align}
from equations~\ref{eq: contrary assumption 2} and~\ref{eq: helper 11}, we have that

\begin{align*}
    C_1(D) + C_1(c^1_2) + C_1(d) \geq C_1(b^1_1 \cup D) + 2C_1(c^1_2),
\end{align*}
which simplifies to 

\begin{align}\label{eq: helper 12}
    C_1(D) + C_1(d) \geq C_1(b^1_1 \cup D) + C_1(c^1_2).
\end{align}

However, since the cost functions are subadditive, we have 

\begin{align}\label{eq: helper 13}
    C_1(D) < C_1(b^1_1 \cup D),
\end{align}
and since $d \notin \{c^1_1, c^2_1\}$, we have

\begin{align}\label{eq: helper 14}
    C_1(d) < C_1(c^1_2).
\end{align}

By combining equations~\ref{eq: helper 13} and~\ref{eq: helper 14}, we have 

\begin{align}\label{eq: helper 15}
    C_1(D) + C_1(d) < C_1(b^1_1 \cup D) + C_1(c^1_2).
\end{align}

Clearly, equation~\ref{eq: helper 12} and~\ref{eq: helper 15} contradict each other, and we have shown that we have $C_1(D) \leq 2C_1(c^1_2)$, which concludes the proof of this lemma.
\end{proof}

We now have

\begin{align*}
    \max_{c \in X_1} C_1(X_1 \setminus c) &= C_1(D) \leq 2C_1(c^1_2) = 2C_1(X_3) \\
    &\leq 2C_1(c^2_2) \leq 2C_1(b^2_1 \cup c^2_2) = 2C_1(X_2),
\end{align*}
where the first inequality comes from Lemma~\ref{lemma: helper 2}, the second inequality comes from the fact that $c^2_2$ and $c^1_2$ are respectively the highest and second highest valued items with respect to $C_1$ and the third inequality comes from the monotonicity of the cost functions. Therefore, agent 1 does not strongly envy agents 2 or 3, thus, property (1) holds in this case.

We now consider the case where agent 2 does not envy agent 1, but strongly envies agent 3, i.e., 

\begin{align}\label{eq: helper 16}
    C_2(b^1_1 \cup D) < C_2(b^2_1 \cup c^1_2),
\end{align}
and

\begin{align}\label{eq: helper 17}
    C_2(X_2) = C_2(b^1_1 \cup D) \geq \max_{c \in X_2}C_2(X_2 \setminus c) > 2C_2(c^2_2).
\end{align}

We make two case arguments for when $C_3(D) \leq C_3(c^2_3)$ and $C_3(D) > C_3(c^2_3)$. Assume that 

\begin{align}\label{eq: assumption 4}
    C_3(D) \leq C_3(c^2_3).
\end{align}

Consider the allocation below

\begin{align*}
    X = \langle \{c^1_2, c^1_3\}, \{c^2_2, c^2_3\}, D \rangle,
\end{align*}
property (2) holds by Lemma~\ref{costly are split}. We have 

\begin{align*}
    C_3(X_3) = C_3(D) \leq C_3(c^2_3) \leq C_3(c^1_3) \leq C_3(c^1_2 \cup c^1_3) = C_3(X_1),
\end{align*}
where the first inequality comes from equation~\ref{eq: assumption 4}, the second inequality comes from the fact that $c^1_3$ and $c^2_3$ are respectively the highest and second highest most costly chore with respect to $C_3$, and the third inequality comes from the monotonicity of the cost functions. Similarly, we have

\begin{align*}
    C_3(X_3) = C_3(D) \leq C_3(c^2_3) \leq C_3(c^2_2 \cup c^2_3) = C_3(X_2),
\end{align*}
where the first inequality comes from equation~\ref{eq: assumption 4} and the second inequality comes from the monotonicity of the cost functions. Therefore, agent 3 does not envy agents 1 or 2. We now argue that agents 1 and 2 do not strongly envy each other. One can observe that 

\begin{align*}
    \max_{c \in X_1}C_1(X_1 \setminus c) = C_1(c^1_2) \leq C_1(c^2_2) \leq C_1(c^2_2 \cup c^2_3) = C_2(X_2),
\end{align*}
where the first inequality comes from the fact that $c^2_2$ and $c^2_1$ are respectively the highest and second highest valued item with respect to $C_1$ and the second inequality comes from the monotonicity of the cost functions. Similarly, we have

\begin{align*}
    \max_{c \in X_2}C_2(X_2 \setminus c) = C_2(c^2_2) \leq C_2(c^1_2) \leq C_2(c^1_2 \cup c^1_3) = C_2(X_1),
\end{align*}
where the first inequality comes from the fact that $c^1_2$ and $c^2_2$ are respectively the highest and second highest valued item with respect to $C_2$, and the second inequality comes from the monotonicity of the cost functions. Therefore, we only have to show that Agents 1 and 2 do not strongly envy agent 3, i.e., for $i \in [2]$, we will show that 

\begin{align*}
    \max_{c \in X_i} C_i(X_i \setminus c) \leq 2C_i(D) = 2C_i(X_3)
\end{align*}

\begin{lemma}\label{lemma: helper 3}
    
Assuming $m \geq 6$, the conditions of case B2222, equations~\ref{eq: helper 8},~\ref{eq: property D2} and~\ref{eq: helper 17} holds, we will have 

    \begin{align*}
        \max_{c \in X_1} C_1(X_1 \setminus c) \leq 2C_1(D) = 2C_1(X_3),
    \end{align*}
    and

    \begin{align*}
        \max_{c \in X_2} C_2(X_2 \setminus c) \leq 2C_2(D) = 2C_2(X_3), 
    \end{align*}
    i.e., agents 1 and 2 will not strongly envy agent 3.
\end{lemma}

\begin{proof}

We will first show that 

\begin{align*}
        \max_{c \in X_1} C_1(X_1 \setminus c) \leq 2C_1(D).
    \end{align*}

Assume the contrary that 

\begin{align*}
        \max_{c \in X_1} C_1(X_1 \setminus c) > 2C_1(D),
    \end{align*}
    and note that $\max_{c \in X_1} C_1(X_1 \setminus c) = C_1(c^1_2)$, therefore, we have

\begin{align}\label{eq: contrary assumption 3}
        C_1(c^1_2) > 2C_1(D),
    \end{align}
therefore, we conclude that 

\begin{align}
        C_1(c^1_2) + C_1(c^1_2) &>  2C_1(D) + 2C_1(b^1_1) \nonumber \\
        &> 2C_1(b^1_1 \cup D) > 2C_1(c^1_2),
    \end{align}
where the first inequality comes from combining equations~\ref{eq: contrary assumption 3} and~\ref{eq: helper 8}, the second inequality comes from the subadditivity of the cost functions, and the third inequality comes from equation~\ref{eq: property D2}. by this contradiction, we conclude that 

\begin{align*}
        \max_{c \in X_1} C_1(X_1 \setminus c) \leq 2C_1(D).
    \end{align*}

we now show that 

\begin{align*}
        \max_{c \in X_2} C_2(X_2 \setminus c) \leq 2C_2(D).
    \end{align*}

    We again assume the contrary that 

    \begin{align*}
        \max_{c \in X_2} C_2(X_2 \setminus c) > 2C_2(D),
    \end{align*}
    and note that $\max_{c \in X_2} C_2(X_2 \setminus c) = C_2(c^2_2)$, therefore we have

    \begin{align}\label{eq: contrary assumption 4}
        C_2(c^2_2) > 2C_2(D),
    \end{align}
    and by combining equations~\ref{eq: helper 17} and~\ref{eq: contrary assumption 4} we have 

    \begin{align*}
        C_2(b^1_1 \cup D) > 4C_2(D),
    \end{align*}
    and from the subadditivity of the cost functions, we have

    \begin{align*}
        C_2(b^1_1) + C_2(D) \geq C_2(b^1_1 \cup D) > 4C_2(D),
    \end{align*}
    which simplifies to

    \begin{align}\label{eq: helper 18}
        C_2(b^1_1) > 3C_2(D),
    \end{align}
    by adding $C_2(b^1_1)$ to both sides of equation~\ref{eq: helper 18} and by the subadditivity of the cost functions we have

    \begin{align}\label{eq: helper 19}
        2C_2(b^1_1) > 3C_2(D) + C_2(b^1_1) \ge 2C_2(D) + C_2(b^1_1 \cup D),
    \end{align}
    where we can conclude that 

    \begin{align*}
        2C_2(b^1_1) > 2C_2(D) + C_2(b^1_1 \cup D) > 2C_2(D) + 2C_2(c^2_2)
    \end{align*}

    where the first inequality comes from equation~\ref{eq: helper 19} and the second inequality comes from equation~\ref{eq: helper 17}. We reach a contradiction since $C_2(D) \geq 0$ and $C_2(c^2_2) \geq C_2(b^1_1)$, i.e., $b^1_1 \notin \{c^1_2, c^2_2\}$. Therefore, we have 

\begin{align*}
        \max_{c \in X_1} C_2(X_2 \setminus c) \leq 2C_2(D),
    \end{align*}
which concludes the proof of this lemma.
\end{proof}

By Lemma~\ref{lemma: helper 3}, we conclude that agents 1 and 2 do not strongly envy agent 3; therefore, property (1) holds in this case.

We now consider the case where 

\begin{align}\label{eq: assumption 5}
    C_3(D) > C_3(c^2_3).
\end{align}

Consider the allocation below

\begin{align*}
    X = \langle D, \{c^2_2, c^1_3\}, \{c^1_2\} \rangle, 
\end{align*}
note that since equation~\ref{eq: assumption 5} holds, property (2) holds with respect to $C_3$, and by Lemma~\ref{costly are split}, property (2) also holds with respect to $C_1$ and $C_2$. Therefore, property (2) will hold. We observe that agent 3 is a singleton. Thus, it will not strongly envy agents 1 and 2. We have that

\begin{align*}
    \max_{c \in X_2}C_2(X_2 \setminus c) = C_2(c^2_2) \leq C_2(c^1_2) ,
\end{align*}
where the inequality comes from the fact that $c^1_2$ and $c^2_2$ are respectively the highest and second highest costly chore with respect to $C_2$. Additionally, we have

\begin{align*}
    \max_{c \in X_2}C_2(X_2 \setminus c) = C_2(c^2_2) \leq 2C_2(D) = 2C_2(X_1),
\end{align*}
where the inequality comes from Lemma~\ref{lemma: helper 3}. Therefore, agent 2 does not strongly envy agents 1 and 3. We now show that agent 1 does not strongly envy agents 2 and 3. One can observe that for any item $d \in D$ we have

\begin{align*}
    C_1(D \setminus d) &\leq C_1(b^1_1 \cup (D \setminus d)) \leq C_1(c^1_2) = C_1(X_3) \\
    &\leq C_1(c^2_2) \leq C_1(c^2_2 \cup c^1_3) = C_1(X_2),
\end{align*}
where the first and fourth inequalities comes from the monotonicity of the cost functions, the second inequality comes from equation~\ref{eq: property D2} and the third inequality comes from the fact that $c^2_2$ and $c^1_2$ are respectively the highest and second highest costly chore with respect to $C_1$. Therefore, agent 1 will not strongly envy agents 2 and 3; thus, property (1) holds.

Now we handle the case where equation~\ref{eq: assumption 3} does not hold, i.e., there is no such subset of items $D \in M'$, or equivalently,

\begin{align}\label{eq: assumption 6}
    C_1(b^1_1 \cup M') < C_1(c^1_2).
\end{align}

Consider the allocation below

\begin{align*}
    X = \langle \{b^1_1\} \cup M', \{c^2_2, b^2_1\}, \{c^1_2\} \rangle,
\end{align*}
we can observe that agent 3 is a singleton and does not strongly envy any other agent. Additionally, we have

\begin{align*}
    C_1(X_1) &= C_1(b^1_1 \cup M') \leq C_1(c^1_2) = C_1(X_3) \\
    &\leq C_1(c^2_2) \leq C_1(c^2_2 \cup b^2_1) = C_1(X_2),
\end{align*}
where the first inequality comes from equation~\ref{eq: assumption 6}, the second inequality comes from the fact that $c^2_2$ and $c^1_2$ are respectively the highest and second highest costly chore with respect to $C_1$ and the third inequality comes from the monotonicity of the cost functions. Therefore, we conclude that agent 1 does not strongly envy agent 2 or 3. We can also see that 

\begin{align*}
    \max_{c \in X_2} C_2(X_2 \setminus c) = C_2(c^2_2) \leq C_2(c^1_2) = C_2(X_3),
\end{align*}
where the inequality comes from the fact that $c^1_2$ and $c^2_2$ are respectively the highest and second highest costly chore with respect to $C_2$, thus agent 2 does not strongly envy agent 3. If agent 2 does not strongly envy agent 1, we can return a full 2-EFX allocation. Therefore, assume agent 2 strongly envies agent 1, i.e., 

\begin{align}\label{eq: assumption 7}
    \max_{c \in X_2}C_2(X_2 \setminus c) = C_2(c^2_2) > 2C_2(b^1_1 \cup M'),
\end{align}
we now consider the allocation below

\begin{align*}
    X = \langle\{c^2_2\}, \{b^1_1, b^2_1\} \cup M', \{c^2_1\} \rangle,
\end{align*}
note that agents 1 and 3 are singleton and do not strongly envy any other agents. We can observe that

\begin{align*}
    C_2(b^1_1 \cup b^2_1 \cup M') &\leq C_2(b^1_1 \cup M') + C_2(b^2_1) \leq C_2(c^2_2) + C_2(b^2_1) \\
    &\leq 2C_2(c^2_2) = 2C_2(X_1) \leq 2C_2(c^1_2) = 2C_2(X_3). 
\end{align*}
where the first inequality comes from the subadditivity of the cost functions, the second inequality comes from equation~\ref{eq: assumption 7}, the third inequality comes from the fact that $b^2_1 \notin \{c^1_2, c^2_2\}$ and the fourth inequality comes from the fact that $c^1_2$, $c^2_2$ are respectively the highest and second highest costly chore with respect to $C_2$. Therefore, agent 2 does not strongly envy agents 1 and 3, and we will return a full 2-EFX allocation. Therefore, in all the cases, we will either return a partial EFX allocation satisfying properties (1) and (2) or return a full 2-EFX allocation.

\paragraph{Case C, $\abs{\sset{c^1_1,c^1_2,c^1_3}} = 2$ and not case B:} W.l.o.g assume $c^1_1 = c^1_2 \neq c^1_3$ and $c^2_1 \neq c^2_2$. Allocate $c^1_1$ to agent 3, $c^2_1$ to agent 2, and $c^2_2$ to agent 1 (we can do so since these are distinct items). We then let agents 1 and 2 choose the most costly remaining chore with respect to $C_3$ (in an arbitrary order). Agent 3 has a single item and, therefore, does not strongly envy any other agent. Agents 1 and 2 have two items. They do not strongly envy any other agents since their two most costly chores are distributed among the other two agents. Therefore, property (1) holds. Property (2) holds by Lemma~\ref{costly are split}. 

\paragraph{Case D, $\abs{\sset{c^{1}_1, c^{1}_2, c^{1}_3}} = 3$ and not case B:} We divide the problem into sub-cases based on the second most costly chores.

\paragraph{Case D1, There exists two agents $i, j$ such that $c^{1}_i = c^{2}_j$:} In this case w.l.o.g assume that $c^{1}_1 = c^{2}_2$. By the case definition, we must have that $c^{2}_1 \ne c^{1}_2$, and $c^{1}_3$ is not equal to $c^1_2$. Allocate $c^1_1$ to agent 3, $c^1_2$ to agent 1, and $c^2_1$ to agent 2. We then let agents 1 and 2 choose the most costly remaining chore with respect to $C_3$ (in an arbitrary order). The two most costly chores with respect to any agent are allocated to two distinct bundles, so property (2) holds. Agent 3 has only a single item, and therefore, it does not strongly envy other agents. Agents 1 and 2 have at most two chores, and since any other agents other than $i \in [2]$ has one of the two most costly chores with respect to $C_i$, agent $i$ will not strongly envy any other agents. Therefore, property (1) holds.

\paragraph{Case D2, (= not Case D1), i.e., for any two agents $i, j$ we have  $c^{1}_i \ne c^{2}_j$:} We split further.

\paragraph{Case D21, Case D2 and $\abs{\sset{c^2_1, c^2_2, c^2_3}} = 3$:} In this case, clearly, we have the two most costly chores with respect to any two agents that are completely distinct. Allocate $c^1_2$, $c^2_3$ to agent 1, $c^1_3$, $c^2_1$ to agent 2, and $c^1_1$, $c^2_2$ to agent 3. For any agent $i \in [3]$, $|X_i| 
= 2$, and since $c^1_i$ and $c^2_i$ are allocated to the other two agents and are in distinct bundles, we have that agent $i$ does not strongly envy any other agent and property (2) clearly holds. Clearly, by the same argument, the allocation is EFX, and property (1) holds. Therefore, both properties hold.

\paragraph{Case D22, Case D2 and $\abs{\sset{c^2_1, c^2_2, c^2_3}} = 1$:} Allocate $c^1_1$ to agent 3, $c^2_1$ to agent 2, and $c^1_3$, $c^1_2$ to agent 1. Since the two most costly chores with respect to $C_1$ are allocated to agents 2 and 3, and agent 1 has two chores, agent 1 will not strongly envy any other agent. Since agents 2 and 3 have only a single item, they will not strongly envy any other agent. Therefore, property (1) holds. Based on the case definition, property (2) clearly holds. 

\paragraph{Case D23, Case D2 and $\abs{\sset{c^2_1, c^2_2, c^2_3}} = 2$:} Assume w.l.o.g that $c^2_1 = c^2_2 \neq c^2_3$.  Allocate $c^2_1$ to agent 3, $c^1_1$ to agent 2, and $c^1_2$ to agent 1. We then let agents 1 and 2 choose the most costly chores with respect to $C_3$ (in an arbitrary order). Agents 1 and 2 have chores more costly than any unallocated chore with respect to $C_3$. The same arguments hold for agents 1 and 3 with respect to $C_2$ and agents 2 and 3 with respect to $C_1$. Therefore, property (2) holds. Agent 3 receives one item and will not strongly envy any other agent. Agents 1 and 2 have two chores that are not the two most costly chores with respect to their cost functions (the 2 most costly chores are allocated to the other agents). Therefore, agents 1 and 2 will not strongly envy any other agent. We conclude that property (1) holds.

One can easily see that the cases and sub-cases above expand over all possible intersections between the top three most costly chores of the three agents. Therefore, we can always find a partial allocation satisfying properties (1) and (2) or a full 2-EFX allocation. Note that a full 2-EFX allocation is only obtained in cases B2221 and B2222.

The runtime needed to find an allocation satisfying properties (1) and (2) is polynomial since we only need to sort the chores based on the cost functions and find the corresponding case and sub-case to those cost functions. Moreover, computing a full 2-EFX allocation in cases B2221 and B2222 can be done in polynomial time by Lemmas ~\ref{lemma: finding subset D} and~\ref{lemma: finding subset D2}. Thus, the runtime to either find an allocation satisfying properties (1) and (2) or a full 2-EFX allocation is polynomial. Therefore, by Lemma~\ref{max(alpha, (beta + 1))-EFX chores}, we can always compute a 2-EFX allocation in polynomial time, which concludes the proof of Theorem~\ref{2-EFX for three agents}.

\subsection{Counterexample to~\cite{ijcai2024p0300}'s Algorithm for 2-EFX Allocations with Three Additive Agents}\label{sec: counter example}
In this section, we provide an example where the algorithm in~\cite{ijcai2024p0300} that claims to achieve a 2-EFX allocation for three agents with additive cost functions will not return a 2-EFX allocation. The example will have six chores. 


    
    

\begin{figure}[t!]
    \centering 
  \begin{minipage}[t]{3.5cm}
  \centering 
  \begin{tabular}{|c|c|c|c|}
    
    \hline
     & $C_1$ & $C_2$ & $C_3$  \\ \hline
    $c_1$ & 10 & 6 & 1 \\ \hline
    $c_2$ & 6 & 10 & 3  \\ \hline
    $c_3$ & 4 & 3 & 4  \\ \hline
    $c_4$ & 0.5 & 2 & $\frac{m_1}{2}$  \\ \hline
    $c_5$ & 0.5 & 2 & $\frac{m_1}{2}$  \\ \hline
    $c_6$ & 3 & 1.5 & $m_2$  \\ \hline
\end{tabular}\vspace{1em}

(a)
\end{minipage}\quad
\begin{minipage}[t]{4.7cm}
\centering 
\begin{tabular}{|c|c|c|c|}
    \hline
     & $C_1$ & $C_2$ & $C_3$  \\ \hline
    $X_1 = \sset{c_2}$ & 6 & 10 & 3 \\ \hline
    $X_2 = \sset{c_3}$ & 4 & 3 & 4  \\ \hline
    $X_3 = \sset{c_1}$ & 10 & 6 & 1  \\ \hline
\end{tabular}\vspace{1em}

\begin{tikzpicture}[node distance=1.5cm, auto]

            \node[draw, circle] (1) {1};
            \node[draw, circle, right of=1] (2) {2};
            \node[draw, circle, right of=2] (3) {3};
 
            \draw[->] (1) to (2);

\end{tikzpicture}

(b)
\end{minipage}\quad
\begin{minipage}[t]{4.7cm}
\centering 
\begin{tabular}{|c|c|c|c|} \hline
     & $C_1$ & $C_2$ & $C_3$  \\ \hline
    $X_1 = \sset{c_2}$ & 6 & 10 & 3 \\ \hline
    $X_2 = \sset{c_3,c_4}$ & 4.5 & 5 & $\frac{m_1}{2} + 4$ \\ \hline
    $X_3 = \sset{c_1}$ & 10 & 6 & 1  \\ \hline
\end{tabular}\vspace{1em}

 \begin{tikzpicture}[node distance=1.5cm, auto]

            \node[draw, circle] (1) {1};
            \node[draw, circle, right of=1] (2) {2};
            \node[draw, circle, right of=2] (3) {3};

            \draw[->] (1) to (2);

        \end{tikzpicture}

(c)
\end{minipage}\vspace{1em} \newline

\begin{minipage}[t]{6cm}
\centering

\begin{tabular}{|c|c|c|c|} \hline
     & $C_1$ & $C_2$ & $C_3$  \\ \hline
    $X_1 = \sset{c_2}$ & 6 & 10 & 3 \\ \hline
    $X_2 = \sset{c_3,c_4, c_5}$ & 5 & 7 & $m_1 + 4$ \\ \hline
    $X_3 = \sset{c_1}$ & 10 & 6 & 1  \\ \hline
\end{tabular}\vspace{1.2cm}

 \begin{tikzpicture}[node distance=1.5cm, auto]

            \node[draw, circle] (1) {1};
            \node[draw, circle, right of=1] (2) {2};
            \node[draw, circle, right of=2] (3) {3};

            \draw[->] (1) to (2);
            \draw[->] (2) to (3);

        \end{tikzpicture}

        (d) 
   
\end{minipage}\quad
\hspace{1cm}
\begin{minipage}[t]{6cm}
\centering

\begin{tabular}{|c|c|c|c|} \hline
     & $C_1$ & $C_2$ & $C_3$  \\ \hline
    $X_1 = \sset{c_2}$ & 6 & 10 & 3 \\ \hline
    $X_2 = \sset{c_3,c_4, c_5}$ & 5 & 7 & $m_1 + 4$  \\ \hline
    $X_3 = \sset{c_1,c_6}$ & 13 & 7.5 & $m_2 + 1$  \\ \hline
\end{tabular}\vspace{1em}

        \begin{tikzpicture}[node distance=1.5cm, auto]

            \node[draw, circle] (1) {1};
            \node[draw, circle, right of=1] (2) {2};
            \node[draw, circle, right of=2] (3) {3};

            \draw[->] (1) to (2);
            \draw[->, bend right=60] (3) to (1);

        \end{tikzpicture}

(e)
\end{minipage}
\caption{(a) shows the cost matrix: $C_i(c_j)$ is shown in column $C_i$ and row $c_j$. $m_1$ and $m_2$ are any reals such that $\frac{m_1}{2} > m_2 > 6$. \\ \protect
(b) shows the initial allocation $X_1 = \sset{c_2}$, $X_2 = \sset{c_3}$ and $X_3 = \sset{c_1}$. $C_i(X_j)$ is shown in column $i$ and row $j$. The corresponding top trading envy graph is shown at the bottom. \\ \protect
(c) shows the situation after allocating $c_4$ to agent 2, i.e., it shows the cost of each agent for each bundle and the corresponding top trading envy graph for the partial allocation $X_1 = \{c_2\}$, $X_2 = \{c_3, c_4\}$ and $X_3 = \{c_1\}$. \\ \protect
(d) shows the situation after allocating $c_5$ to agent 2, i.e., it shows the cost of each agent for each bundle and the corresponding top trading envy graph for the partial allocation $X_1 = \{c_2\}$, $X_2 = \{c_3, c_4, c_5\}$ and $X_3 = \{c_1\}$. \\ \protect
(e) shows the situation after allocating $c_6$ to agent 3, i.e., it shows the cost of each agent for each bundle and the corresponding top trading envy graph for the final allocation $X_1 = \{c_2\}$, $X_2 = \{c_3, c_4, c_5\}$ and $X_3 = \{c_1, c_6\}$. }
\label{fig:new_figure}
\end{figure}
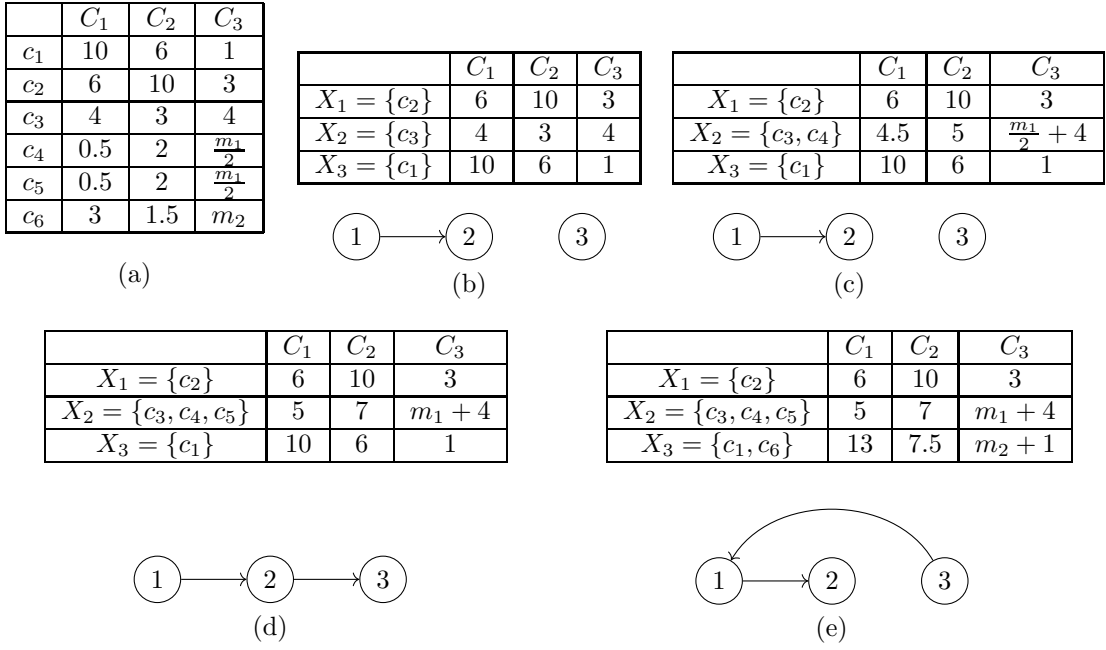

\begin{lemma}
    Consider the setting with three agents with additive cost functions and 6 items given by Table (a) in Figure~\ref{fig:new_figure}, where $m_1$ and $m_2$ are any real numbers such that $\frac{m_1}{2} > m_2 > 6$. Let $X = \{X_1, X_2, X_3\}$ be the output of the algorithm in~\cite{ijcai2024p0300}. Then
    \begin{align*}
        \max_{c \in X_3} C_3(X_3 \setminus c) > 2C_3(X_1),
    \end{align*}
i.e., the algorithm does not construct a 2-EFX allocation. Moreover, the algorithm does not guarantee any constant approximation factor.\end{lemma}

\begin{proof}
    We first note that their approach is similar to ours in the sense that they want to construct
    \begin{enumerate}
        \item[(1)] a partial 2-EFX allocation $X = \{X_1, X_2, X_3\}$ such that \item[(2)] for any agent $i \in [3]$, and any unallocated item $b$, $C_i(b) \leq C_i(X_j)$ for at least two agents $j \in [3]$.
\end{enumerate}
    
    They first distinguish cases whether the top-2 (the 2 most costly chores) of the agents have a common item or not. Clearly, in our case, the top-2 items for agents 1 and 2 have common chores, namely, $c_1$ and $c_2$, and notably, the top-2 items of agent 3 are distinct from the top-2 items of agents 1 and 2 since $c_4$ and $c_5$ are the highest valued items with respect to $C_3$. They next distinguish whether the second highest valued chores of two agents are the same or not. In our case, the second highest valued chores with respect to the three agents are distinct: $c_2$ for agent 1, $c_1$ for agent 2, and either of $c_4$ and $c_5$ for agent 3. Therefore, we are in the case where the highest valued item with respect to some agent belongs to the top-2 items of another agent, namely, $c_1$, which is the highest valued item with respect to agent 1, belongs to the two most costly chores with respect to agent 2, which are $c_1$ and $c_2$. They then construct a partial EFX allocation as follows:

    \begin{enumerate}
        \item allocate $c_1$ to agent 3,
        \item allocate to agent 1 the highest valued unallocated item with respect to $C_2$, i.e, allocate $c_2$ to agent 1,
        \item allocate to agent 2 the highest valued unallocated item with respect to $C_1$, i.e., allocate $c_3$ to agent 2.
    \end{enumerate}

    This is clearly a partial EFX allocation since each agent has exactly one item. Property (2) holds for any unallocated item $b$ and any agent $i \in [2]$. However, it does not hold for agent 3 since $c_4$ and $c_5$, which are the highest valued items with respect to $C_3$, are unallocated. They claim that by running the top trading envy cycle elimination algorithm with respect to $C_3$, i.e., at each step allocating the chore with the highest value with respect to $C_3$, and choosing the sinks in a lexicographic manner, i.e., giving priority to agent 1, then agent 2, and then agent 3, before allocating any item to agent 3, i.e., agent 3 becoming a sink, we will allocate at least one item to the rest of the agents, i.e., agents 1 and 2 will become a sink before agent 3. We show that this is not the case. 
    
    
    The values of the bundles of the initial allocation for the three agents and the corresponding top trading envy graph are shown in Figure~\ref{fig:new_figure}(b), i.e., the minimum valued bundle with respect to $C_1$ is $X_2$, and agents 2 and 3 each have their minimum valued bundles with respect to their cost function, so we only have an edge from agent 1 to agent 2. Therefore, the top trading envy cycle elimination algorithm will choose agent 2 as the sink and allocate one of $c_4$ or $c_5$ to agent 2 since $\frac{m_1}{2} > m_2$. Note that since $c_4$ and $c_5$ have the same value with respect to all three agents, we can assume, without loss of generality, that we allocate $c_4$ to agent 2 in this step.

    The values of the bundles of the initial allocation for the three agents and the corresponding top trading envy graph are shown in Figure~\ref{fig:new_figure}(c). One can observe that the top trading envy graph will not change after allocating $c_4$ to agent 2 in the previous step. Therefore, the top trading envy cycle elimination algorithm will choose agent 2 as the sink and allocate $c_5$ to agent 2 since $\frac{m_1}{2} > m_2$.
    
    The values of the bundles and the corresponding top trading envy graph are now shown in Figure~\ref{fig:new_figure}(d). One can observe that the minimum valued bundles with respect to $C_1$ and $C_2$ are $X_2$ and $X_3$, respectively, and agent 3 has its minimum valued bundle with respect to its cost function. Therefore, agent 3 is the only sink, and the algorithm allocates $c_6$ to agent 3.

    Finally, the values of the bundles and the corresponding top trading envy graph for the final allocation are shown in Figure~\ref{fig:new_figure}(e). We can see that the minimum valued bundle with respect to $C_3$ and $C_1$ are $X_1$ and $X_2$, respectively, and agent 2 has its minimum valued bundle with respect to its cost function. Thus, there is no cycle in the top trading envy graph, and hence the algorithm will terminate with the allocation $X_1 = \{c_2\}$,  $X_2 = \{c_3, c_4, c_5\}$ and $X_3 = \{c_1, c_6\}$. However, 
    
\begin{align*}\max_{c \in X_3} C_3(X_3 \setminus c)& = C_3(X_3 \setminus c_1) = C_3(c_6) = m_2 \\
&> 6 = 2C_3(c_2) = 2C_3(X_1), \end{align*}
and therefore, the allocation is not 2-EFX. Moreover, 
\[
        \frac{\max_{c \in X_3} C_3(X_3 \setminus c)}{C_3(X_1)} = \frac{m_2}{3},
        \]
         and hence by increasing the values of $m_1$ and $m_2$ such that $\frac{m_1}{2} > m_2$, no approximation guarantee can be given for this algorithm in this case, concluding the proof of the lemma. 

\end{proof}

We now demonstrate how our algorithm (Theorem~\ref{2-EFX for three agents}) behaves in this setting. Note that since we have that the highest valued item with respect to each agent are distinct items, and there exists two agents, namely agents 1 and 2, who agree on the two most costly chores with respect to their cost functions, the algorithm will be at case B. Furthermore since we have that the two most costly chores with respect to agent 3 and the two most costly chores with respect to agent 1 do not have common items, the algorithm will be at case B2. We now note that agents 1 and 2 agree on the third most costly chore, and from the fact that this chore does not belong among the two most costly chore with respect to $C_3$, and since the most costly chore with respect to agent 1 is the second most costly chore with respect to agent 2 and visa versa, the algorithm will be at case B2222. 

First note that since the two most costly chores with respect to $C_3$ are identical items, we can wlog assume that $b^1_1 = b^1_2 = c_4$ and $b^2_1 = b^2_2 = c_5$.  We first construct a partial allocation as below

\begin{enumerate}
    \item allocate $c^2_2$, i.e., $c_1$ to agent 3,
    \item allocate $b^2_1$ and $c^1_2$, i.e., $c_5$ and $c_2$ to agent 1,
    \item allocate $b^1_1$, i.e., $c_4$ to agent 2.
\end{enumerate}

At this point, property (2) clearly holds by Lemma~\ref{costly are split}. The only strong envy existing in this partial allocation is from agent 1 to agent 2, since we have  

\begin{align*}
    \max_{c \in X_1} C_1(X_1 \setminus c) = C_1(c_2) = 6 > 1 = 2C_1(c_4) = 2C_1(X_2),
\end{align*}
therefore, since we have

\begin{align*}
    C_1(\{c_4, c_3, c_6\}) = 7.5 > 6 \geq C_1(c^2_1),
\end{align*}
or equivalently, equation~\ref{eq: assumption 3} holds, by Lemma~\ref{lemma: finding subset D2}, we can find a non-empty subset of unallocated items $D$ such that equation~\ref{eq: property D2} holds for any $d \in D$. One can observe that this set $D$ is equal to the set of unallocated items, i.e., $D = \{c_3, c_6\}$. We allocate this set to agent 2. We can now see that our allocation is a 2-EFX allocation. Agent 3 is singleton and thus does not strongly envy any other agent. Agent 1 does not envy any other agent since 

\begin{align*}
    C_1(X_1) = 6.5 < 7.5 = C_1(X_2) < 10 = C_1(X_3),
\end{align*}
and similarly, agent 2 does not strongly envy any other agent since 

\begin{align*}
    \max_{c \in X_2} C_2(X_2 \setminus c) = C_2(\{c_4, c_5\}) = 5 < 6 = C_2(X_3) < 12 = C_2(X_3),
\end{align*}
thus, our algorithm returns the allocation $X_1 = \{c_2, c_5\}$, $X_2 = \{c_3, c_4, c_6\}$ and $X_3 = \{c_1\}$ which is a 2-EFX allocation (it is also an EFX allocation).

\section{tEFX Existence with Three Groups of Agents}\label{tEFX Existence for a General and 2-ratio-bounded Cost Functions}
  
In this section, we show that a tEFX allocation exists for three agents when one agent has an additive 2-ratio-bounded cost function.

\begin{theorem} \label{3 restricted agents tEFX no group} A tEFX allocation is guaranteed to exist for three agents if one agent has an additive 2-ratio-bounded valuation function and the other two have general cost functions. 
\end{theorem}

We provide a more general result where a tEFX allocation exists when the agents can be categorized into three groups based on their cost functions: one group with general monotone cost functions $C_1(\cdot)$, one group with additive and 2-ratio-bounded cost functions $C_2(\cdot)$, and one group consisting of one agent with a general monotone cost function $C_3(\cdot)$. 

\begin{theorem} \label{3 restricted agents tEFX}
A tEFX allocation of chores exists in a setting where agents are divided into three distinct groups based on their cost function types. The groups include agents with identical general monotone cost functions, agents with identical additive 2-ratio-bounded cost functions, and a single agent with a general monotone cost function. 
\end{theorem}

We recall that by~\cite{ijcai2023p0277}, it is sufficient to find tEFX allocations for non-degenerate instances. Thus, from now on, we assume our cost functions are non-degenerate. Our proof relies on the following lemma, which shows the existence of a tEFX allocation with additional properties when the agents are split into two groups based on their cost functions; see Algorithm~\ref{tEFX two groups}.

\begin{lemma} \label{tEFX lemma}
    If the agents are split into two non-empty groups based on their cost functions, one group with identical general cost functions $C_1$, and one group with identical additive and 2-ratio-bounded cost functions $C_2$, then given access to a value oracle for the general monotone cost functions, one can construct an allocation $X = (X_1, \ldots, X_n)$ satisfying the following properties: 
    \begin{itemize}
        \item $\langle X_1, \ldots, X_{n-k+1}\rangle$ are EFX feasible w.r.t $C_1$
        \item $\langle X_{n-k+1}, \ldots, X_{n}\rangle$ are tEFX feasible w.r.t $C_2$
    \end{itemize}
    where $k$ is the number of agents with cost function $C_2$. 
\end{lemma}

\begin{proof}
    We use induction on $k$. If $k = 1$, we run the standard PR\footnote{By the PR algorithm, we mean the algorithm that computes an EFX allocation for agents with identical cost functions.} \cite{doi:10.1137/19M124397X} algorithm for $n$ agents with cost functions $C_1$ and let $X_n$ be the least costly bundle w.r.t cost function $C_2$. Clearly, the resulting allocation satisfies our desired properties.

    For the inductive step, we construct a solution for $k$ from a solution for $k - 1$ by an iterative algorithm that maintains the two invariants below:
    
    \begin{enumerate}
        \item $\langle X_1,\ldots ,X_{n-k+1}\rangle $ be EFX-feasible w.r.t $C_1$.
        \item $\langle X_{n-k+2},\ldots ,X_{n}\rangle $ be tEFX-feasible w.r.t $C_2$,
    \end{enumerate}
    
    and decreases the potential function
    
     \[ \Phi(X) = \sum_{i = 1}^{n-k+1} |X_i| .\]

      Eventually, reaching an allocation satisfying:

    \begin{itemize}
        \item $\langle X_1,\ldots ,X_{n-k+1}\rangle $ be EFX-feasible w.r.t $C_1$.
        \item $\langle X_{n-k+1},\ldots ,X_{n}\rangle $ be tEFX-feasible w.r.t $C_2$,
    \end{itemize}

    which will conclude our proof.
     
    Let $X$ be a solution to the problem for $k-1$. Note that $X$ satisfies the invariants. For each $i \in [n]$, let 
        $d_i = \argmin_{c \in X_i} C_2(c)$.
    
If $X_i$ is tEFX-feasible w.r.t $C_2$ for some $i \in [n-k+1]$, we return $X$. So assume otherwise, i.e., for all $i \in [n-k + 1]$ we have $C_2(X_i \setminus  d_i) > \min_{n-k + 2\leq j \leq n} C_2(X_j \cup d_i)$. We can assume that $C_2(X_n) = \min_{n-k + 2\leq j \leq n} C_2(X_j) = \min_{j \in [n]} C_2(X_j)$ where the second equality comes from the fact that $X_i$ for $i \in [n-k + 1]$ are not tEFX-feasible w.r.t $C_2$. We assume that $C_1(X_1 \setminus  c_1') = \max_{i \in [n-k+1]} \max_{c \in X_i} C_1(X_i\setminus c)$ and define a new allocation $X' = \langle X_1\setminus c_1', X_2,\ldots ,X_{n-1}, X_{n} \cup c_1'\rangle $. We claim that the new allocation satisfies the invariants, and the potential function has decreased.

\begin{algorithm}[t!]
\caption{tEFX-TwoGroup: Find an allocation satisfying properties in Lemma~\ref{tEFX lemma}} \label{tEFX two groups}
\KwIn{$M : \text{set of chores}$, $n : \text{total number of agents}$, $C_1 : \text{general monotone cost function}$, $C_2 : \text{additive 2-ratio-bounded cost function}$, $k : \text{number of agents with cost function $C_2$}$}
\KwOut{Allocation $X = (X_1, \ldots, X_n)$ satisfying properties in Lemma~\ref{tEFX lemma}}

\If{$k = 1$}{

    $X \gets \text{output of the PR algorithm with cost function $C_1$}$

    w.l.o.g assume $C_2(X_n) = \min_{i \in [n]}C_2(X_i)$

    return $X$
}

$X \gets \text{tEFX-TwoGroup($M$, $n$, $C_1$, $C_2$, $k - 1$)}$

\While{$X \text{ does not satisfy the desired properties}$}{

    w.l.o.g assume $C_2(X_n) = \min_{i \in [n]}C_2(X_i)$

    w.l.o.g assume $C_1(X_1 \setminus c_1') = \max_{i \in [n-k+1]} \max_{c \in X_i} (C_1(X_i\setminus c))$

    $X \gets \langle X_1 \setminus c_1', X_2, \ldots, X_{n-1}, X_n \cup c_1'\rangle$

}

return $X$

\end{algorithm}

    \begin{claim}\label{tEFX invariant 1}
    If $X$ satisfies the invariants , the allocation $X' = \langle X_1\setminus c_1', X_2,\ldots ,X_{n-1}, X_{n} \cup c_1'\rangle $ satisfies invariant 1. 
    \end{claim}
    
    \begin{proof}
    For all $2 \leq i \leq n$ and $c \in X_1'$, $C_1(X_1' \setminus  c) \leq C_1(X_1 \setminus  c) \leq C_1(X_i) \leq C_1(X_i') $ thus $X_1'$ is EFX-feasible w.r.t $C_1$. For all $2\leq i \leq n-k+1$ and $2\leq j \leq n$ and $c \in X_i'$ we have that $C_1(X_i' \setminus  c) = C_1(X_i \setminus  c) \leq C_1(X_j) \leq C_1(X_j') $ and also $C_1(X_i' \setminus  c) \leq C_1(X_1 \setminus  c_1') = C_1(X_1') $ so for all $2\leq i \leq n-k+1$ we have that $X_i'$ is EFX-feasible w.r.t $C_1$ and invariant 1 holds.
    \end{proof}

    \begin{claim}\label{tEFX invariant 2}
    If $X$ satisfies the invariants and for all $i \in [n-k + 1]$, $X_i$ is not tEFX-feasible w.r.t $C_2$, the allocation $X' = \langle X_1\setminus c_1', X_2,\ldots ,X_{n-1}, X_{n} \cup c_1'\rangle $ satisfies invariant 2. 
    \end{claim}
    
    \begin{proof}

    For all $2\leq i \leq n-1$ and $c \in X_n'$ we have that 
    \begin{align*}
        C_2(X_n' \setminus  c) &= C_2((X_n \cup c_1') \setminus  c) \leq C_2(X_n \cup c) \\
        &\leq C_2(X_i \cup c) = C_2(X_i' \cup c)
    \end{align*}
    where the first inequality comes from the fact that $C_2$ is 2-ratio-bounded and $C_2(c_1') \leq 2C_2(c)$ and the second inequality comes from the fact that $C_2(X_n) = \min_{i\in [n]} C_2(X_i)$. Therefore, $X_n'$ does not tEFX-envy any of $X_i'$ for $2\leq i \leq n-1$. We now show that $X_n'$ does not tEFX-envy $X_1'$.
    
    For all $c \in X_n'$ we have that 
    \begin{align*}
        C_2(X_n') - 2C_2(c) &= C_2(X_n) + C_2(c_1') - 2C_2(c) \\
        &< C_2(X_1) - 2C_2(d_1) + C_2(c_1') - 2C_2(c) \\
        &\leq C_2(X_1) - C_2(c_1')
    \end{align*}
    where the first inequality comes from the fact that $C_2(X_n \cup d_1) < C_2(X_1 \setminus d_1)$. The second inequality comes from the fact that $C_2$ is 2-ratio-bounded. Therefore, $X_n'$ does not tEFX-envy $X_1'$. Accordingly, we have shown that $X_n'$ is tEFX-feasible w.r.t $C_2$ in $X'$.

    Now we show that for all $n-k+2\leq i \leq n-1$, $X_i'$ is tEFX-feasible w.r.t $C_2$. For all $2\leq j \leq n$ and $c \in X_i'$, 
    $$C_2(X_i' \setminus  c) = C_2(X_i \setminus  c) \leq C_2(X_j \cup c) \leq C_2(X_j' \cup c).$$ Therefore, $X_i$ does not tEFX-envy $X_j$ w.r.t $C_2$. We now show that for all $n-k+2\leq i \leq n-1$, $X_i'$ does not tEFX-envy $X_1'$ w.r.t $C_2$ in $X'$. We have that 
    \begin{align*}
        C_2(X_i' \setminus  c) &= C_2(X_i \setminus  c) \leq C_2(X_n \cup c) \\ 
        &\leq C_2((X_1 \setminus  c_1') \cup c) = C_2(X_1' \cup c)
    \end{align*}
    where the first inequality comes from the fact that invariant 2 holds for $X$. To show the second inequality, we assume the contrary that $C_2((X_1\setminus c_1') \cup c) < C_2(X_n \cup c)$. Since $C_2$ is additive, we have $C_2(X_1\setminus c_1') < C_2(X_n)$. Moreover, $C_2$ is 2-ratio-bounded and as a result, for all $ c' \in X_1$ and $i \in [n]$, 
    $$C_2(X_1) - 2C_2(c') \leq C_2(X_1) - C_2(c_1') < C_2(X_n) \leq C_2(X_i).$$
    Therefore, we must have that $C_2(X_1 \setminus c') < C_2(X_i \cup c')$ so $X_1$ was tEFX-feasible w.r.t $C_2$ in $X$ and we reached a contradiction. Thus, we have shown that for all $n-k+2\leq i \leq n-1$, $X_i'$ is tEFX-feasible w.r.t $C_2$.
    
    \end{proof}

    It remains to be shown that the potential function has decreased. We have  
    \begin{align*}
        \Phi(X') = \sum_{i = 1}^{n-k+1} |X_i'| &= |X_1 \setminus  c_1'| + \sum_{i = 2}^{n-k+1} |X_i| 
        = \Phi(X) - 1.
    \end{align*}
    Since $\Phi(X) \geq 0$, the algorithm terminates. 
     
\end{proof}

The following corollary is also obtained by Lemma~\ref{tEFX lemma}, which is a minor result.

\begin{algorithm}[t!]
\caption{tEFX allocation when agents split to three groups according to Theorem~\ref{3 restricted agents tEFX}} \label{tEFX three groups}

\KwIn{$M : \text{set of chores}$, $n : \text{total number of agents}$, $C_1, C_2, C_3:$ cost functions as described in section \ref{tEFX Existence for a General and 2-ratio-bounded Cost Functions}, $\ell = \text{number of agents with cost function $C_2$}$}
\KwOut{tEFX allocation for instances described in Theorem~\ref{3 restricted agents tEFX}}

$X \gets \text{tEFX-TwoGroup($M$, $n$, $C_1$, $C_2$, $\ell + 1$)}$

$i \gets \argmin_{i \in [n]} C_3(X_i)$

Allocate $X_i$ to the agent with cost function $C_3$

\If{$i \le n - \ell$}{

Allocate bundles $\{X_1, \ldots X_{n-\ell}\} \setminus \{X_i\}$ to the group with cost function $C_1$

Allocate bundles $\{X_{n-\ell+1}, \ldots, X_n\}$ to the group with cost function $C_2$

}

\Else{

Allocate bundles $\{X_1, \ldots X_{n-\ell-1}\}$ to the group with cost function $C_1$

Allocate bundles $\{X_{n-\ell}, \ldots, X_n\} \setminus \{X_i\}$ to the group with cost function $C_2$

}

Return the allocated bundles for each agent

\end{algorithm}
    
    \begin{corollary}
    If the agents are split into two groups based on their cost functions, one group with identical general cost functions $C_1$, and one group with identical additive and 2-ratio-bounded cost functions $C_2$, then a tEFX allocation exists. 
    \end{corollary}

    Let $\ell$ be the number of agents with cost function $C_2$. We prove Theorem~\ref{3 restricted agents tEFX}, which states that a tEFX allocation exists when the agents can be divided into three groups based on their cost functions (one group with general monotone cost function $C_1$, one group with additive and 2-ratio-bounded cost function $C_2$, and a group consisting of one agent with general monotone cost function $C_3$), based on Algorithm \ref{tEFX three groups}. We first find an allocation $X$ according to Lemma~\ref{tEFX lemma} with $k = \ell + 1$. Therefore, allocation $X$ has the following properties:

    \begin{itemize}
        \item $\langle X_1, \ldots, X_{n-\ell}\rangle$ are tEFX feasible w.r.t $C_1$
        \item $\langle X_{n-\ell}, \ldots, X_{n}\rangle$ are tEFX feasible w.r.t $C_2$
    \end{itemize}
    
    Then, we let the agent in the third group choose its least costly bundle among $X_1,\ldots ,X_{n}$; thereafter we can allocate the rest of the bundles to the remaining agents to reach a tEFX allocation in the following manner: if the least costly chore for the agent in the third group is $X_i$ such that $i \leq n - \ell$, we allocate $X_{n-\ell+1}, \ldots, X_n$ to the group with cost function $C_2$, allocate $\{X_1, \ldots , X_{n-\ell}\} \setminus \{X_i\}$ to the group with cost function $C_1$. If the least costly chore for the agent in the third group is $X_i$ such that $i > n - \ell$, we allocate $X_{1}, \ldots, X_{n-\ell-1}$ to the group with cost function $C_1$, allocate $\{X_{n-\ell}, \ldots , X_{n}\} \setminus \{X_i\}$ to the group with cost function $C_2$. It is clear that such allocations are tEFX. Therefore, we have proved Theorem~\ref{3 restricted agents tEFX}.

\section{Relaxations of EFX for $\alpha$-ratio-bounded Cost Functions} \label{Relations of EFX for alpha-ratio-bounded Cost Functions}

In this section, we will show the existence of relaxations of EFX for $n$ agents with ratio-bounded cost functions. We first show that for $n$ agents with additive $\alpha$-ratio-bounded cost functions, a $(1 + \frac{\alpha - 1}{\ceil{m/n} - 1})$-EFX allocation always exists. 

\begin{theorem} \label{alpha ratio theorem}An $(1 + \frac{\alpha - 1}{\ceil{m/n} - 1})$-EFX allocation exists for any number of agents with additive and $\alpha$-ratio-bounded cost functions. Additionally, it can be computed in polynomial time.\end{theorem}

We next show that a tEFX allocation exists for $n$ agents with additive 2-ratio bounded cost functions. 

\begin{theorem} \label{tEFX 2-ratio}A tEFX allocation exists for any number of agents with additive and 2-ratio-bounded cost functions. Additionally, it can be computed in polynomial time.\end{theorem}

 Since it has been shown in \cite{10.1007/978-3-031-43254-5_15} that EFX allocations can always be computed in polynomial time for additive cost functions when $m \le 2n$, we may assume that $\lceil \frac{m}{n} \rceil \ge 3$. We allocate the chores in a simple round-robin manner. 

\begin{definition}[Round-Robin]
    The round-robin procedure to allocate the chores is as follows: we first fix an ordering of the agents $i_1, ..., i_n$. Then, until all the chores are allocated, we run iterations such that in each of them, from $t = 1, ..., n$, each agent $i_t$ chooses the least costly chore with respect to its cost function among the remaining unallocated items, and we allocate that chore to this agent.
\end{definition}
\begin{lemma} \label{(1 + (alpha - 1)/(l-1)-EFX chores}
The output  of the round-robin algorithm after \(\ell \geq 3\) rounds, is \((1 + \frac{\alpha - 1}{\ell - 1})\)-EFX.
\end{lemma}

\begin{proof}
Let \(X = \{X_1, X_2, \ldots, X_n\}\) be the output of the round-robin algorithm after \(\ell \geq 3\) rounds. Let \(i\) and \(j\) be two arbitrary agents such that \(i\) chooses items before \(j\) in the round-robin algorithm. For \( k \in [n]\) and \(t \in  [\ell]\) let \(c^t_k\), be the item selected by agent \(k\) at round \(t\) of our round-robin algorithm. Note that \(c^l_k\) might be undefined for some \(k\).

We first show that for any $c \in X_i$, we have $C_i(X_i \setminus c) \le (1 + \frac{\alpha - 1}{\ell - 1})C_i(X_j)$. Clearly, we only need to show this for the case where agent $i$ has $\ell$ chores and agent $j$ has $\ell-1$ chores. Since the chores have been chosen in a round-robin manner, we have \(C_k(c^{s}_{k}) \leq C_k(c^{t}_{k})\) for any agent \(k\) and \(s \leq t\). Thus for any \(c \in X_i\) we have \(C_i(X_i \setminus {c}) \leq C_i(X_i \setminus {c^1_{i}}) = \sum_{t = 2}^{\ell}  C_i(c^t_{i}) \). Since agent \(i\) chooses its chores before agent \(j\) in the round-robin algorithm, for any \(t \in [\ell-1]\), we have \(C_i(c^t_{i}) \leq C_i(c^t_{j})\). Also, since the cost functions are \(\alpha\)-ratio-bounded we have \(C_i(c^l_{i}) \leq \alpha C_i(c^1_{j})\), and \(C_i(c^l_{i}) \leq \alpha [\min_{2 \leq s \leq \ell-1}(C_i(c^s_{j}))] \leq \alpha \frac{\sum_{t = 2}^{\ell-1} C_i(c^t_{j})}{\ell-2}\). This implies 
\begin{align}
    C_i(c^l_{i}) \leq  \alpha [\beta C_i(c^1_{j})  + (1-\beta)\frac{\sum_{t = 2}^{\ell-1} C_i(c^t_{j})}{\ell-2}]
\end{align}%
for any parameter \(0 \leq \beta \leq 1\). Therefore, 
\begin{align*}
    C_i(X_i \setminus {c}) &\leq  \sum_{t = 2}^{\ell}  C_i(c^t_{i}) \\ \
    & \leq 
    \alpha \beta C_i(c^1_j) + \sum_{t = 2}^{\ell-1} (1 + \frac{\alpha(1 - \beta)}{\ell - 2}) C_i(c^t_j) 
\end{align*}%
\noindent 
and by setting \(\beta = \frac{\ell + \alpha - 2}{\alpha (\ell-1)}\),
\begin{align}
    C_i(X_i \setminus {c}) &\leq  \sum_{t = 1}^{\ell-1}  C_i(c^t_{j})(\frac{\ell + \alpha - 2}{\ell - 1})  \\ \nonumber 
    & = (1 + \frac{\alpha - 1}{\ell - 1})C_i(X_j).
\end{align}%

Now we show for any $c \in X_j$, we have $C_j(X_j \setminus c) \le (1 + \frac{\alpha - 2}{\ell - 1})C_j(X_i)$. Clearly, we only need to show this for the case where agent $i$ and $j$ both have the same number of chores, i.e., both have either $\ell$ or $\ell-1$ chores. Let $\ell'$ be the number of chores in $i$ and $j$. For any \(c \in X_j\) we have \(C_j(X_j \setminus {c}) \leq C_j(X_j \setminus {c^1_{j}}) = \sum_{t = 2}^{\ell'}  C_j(c^t_{j}) \). Clearly for any \(t \in [\ell'-1]\), we have \(C_j(c^t_{j}) \leq C_j(c^{t+1}_{i})\). Also, since the cost functions are \(\alpha \)-ratio-bounded we have that \(C_j(c^{\ell'}_{j}) \leq \alpha[\min_{s \in [2]}(C_j(c^s_{i}))] \leq \alpha \frac{C_j(c^1_{i}) + C_j(c^2_{i})}{2} \), and \(C_j(c^{\ell'}_{j}) \leq \alpha [\min_{3 \leq s \leq \ell'}(C_j(c^s_{i}))] \leq \alpha \frac{\sum_{t = 3}^{\ell'} C_j(c^t_{i})}{\ell'-2}\). Therefore, for any parameter \(0 \leq \beta \leq 1\), we have
\begin{align}
     C_j(c^{\ell'}_{j}) \leq \alpha [\frac{\beta}{2} [C_j(c^1_{i}) + C_j(c^2_{i})] + (1-\beta)\frac{\sum_{t = 3}^{\ell'} C_j(c^t_{i})}{\ell'-2}]
\end{align}
\noindent 
and by setting \(\beta = \frac{2(\alpha + \ell' - 2)}{\alpha (\ell' - 1)}\) we have 
\begin{align}
     &C_j(X_j \setminus {c}) \leq   \sum_{t = 2}^{\ell'}  C_j(c^t_{j}) \le \sum_{t = 3}^{\ell'}  C_j(c^t_{i}) + C_j(c_j^{\ell'}) \leq  \\ \nonumber & \sum_{t = 1}^{\ell'}  C_j(c^t_{i})(\frac{\alpha + \ell' - 2}{\ell'}) \leq (1 + \frac{\alpha - 2}{\ell-1})C_j(X_i).
\end{align}

Therefore we conclude that \(X\) is a max\((1 + \frac{\alpha - 1}{\ell - 1}, 1 + \frac{\alpha - 2}{\ell - 1})\)-EFX and therefore we have a $(1 + \frac{\alpha - 1}{\ell - 1})$-EFX allocation.
\end{proof}

Applying Lemma~\ref{(1 + (alpha - 1)/(l-1)-EFX chores} for $\ell = \lceil \frac{m}{n} \rceil$ to get a full allocation $X$ that is a $(1 + \frac{\alpha - 1}{\lceil \frac{m}{n} \rceil - 1})$-EFX. 

\begin{lemma} \label{tEFX for three agents lemma}
The output of the round-robin algorithm after \(\ell \geq 3\) rounds is tEFX for agents with additive 2-ratio-bounded cost functions.
\end{lemma}

\begin{proof}
Let \(X = \{X_1, X_2, \ldots, X_n\}\) be the output of the round-robin algorithm after \(\ell \geq 3\) rounds. let \(i\) and \(j\) be two arbitrary agents such that \(i\) chooses items before \(j\) in the round-robin algorithm. For \( k \in [n]\) and \(t \in  [\ell]\) let \(c^t_k\), be the item selected by agent \(k\) at round \(t\) of our round-robin algorithm. Note that \(c^l_k\) might be equal to the empty set for some \(k\). 

We first show that for any $c \in X_i$, we have that $C_i(X_i \setminus c) \le C_i(X_j \cup c)$. Clearly, we only need to show this for the case where agent $i$ has $\ell$ items, and agent $j$ has $\ell-1$ items. Since the chores have been chosen in a round-robin manner, we have \(C_k(c^{s}_{k}) \leq C_k(c^{t}_{k})\) for any agent \(k\) and \(s \leq t\). Thus, for any \(c \in X_i\) we have \(C_i(X_i \setminus {c}) \leq C_i(X_i \setminus {c^1_{i}}) = \sum_{t = 2}^{\ell}  C_i(c^t_{i}) \) and \(C_i(X_j \cup {c^1_i}) \leq C_i(X_j \cup {c}) \), therefore we only need to show that $C_i(X_i \setminus c^1_i) \le C_i(X_j \cup c^1_i)$. Since agent \(i\) chooses its chores before agent \(j\) in the round-robin algorithm, for any \(t \in [\ell-1]\), we have \(C_i(c^t_{i}) \leq C_i(c^t_{j})\). Also, since the cost functions are 2-ratio-bounded we have that \(C_i(c^l_{i}) \leq 2 [\min(C_i(c^1_j), C_i(c^1_i))] \leq C_i(c^1_j) + C_i(c^1_i)\), therefore by combining the inequalities we have $\sum_{t = 2}^{\ell} C_i(c^t_i) \le C_i(c^1_i) + \sum_{t = 1}^{\ell - 1} C_i(c^t_j)$ and agent $i$ does not strongly envy agent $j$.

We next show that for any $c \in X_j$, we have that $C_j(X_j \setminus c) \le C_j(X_i \cup c)$. We only need to show this for the case where agent $i$ and $j$ have the same number of items. Let $\ell'$ be the number of items allocated to $i$  and $j$. Since the chores have been chosen in a round-robin manner, we have \(C_k(c^s_{k}) \leq C_k(c^{t}_{k})\) for any agent \(k\) and \(s \leq t\), therefore for any \(c \in X_j\) we have \(C_j(X_j \setminus {c}) \leq C_j(X_j \setminus {c^1_{j}}) = \sum_{t = 2}^{\ell'}  C_j(c^t_{j}) \) and \(C_j(X_i \cup {c^1_j}) \leq C_j(X_i \cup {c}) \), therefore we only need to show that $C_j(X_j \setminus c^1_j) \le C_j(X_i \cup c^1_j)$. Clearly, for any \(t \in [\ell'-1]\), we have \(C_j(c^t_{j}) \leq C_j(c^{t+1}_{i})\). Also, since the cost functions are 2-ratio-bounded we have that \(C_j(c^{\ell'}_{j}) \leq 2 C_j(c^1_{i})\), and \(C_j(c^{\ell'}_{j}) \leq 2 C_j(c^2_{i})\), therefore, \(C_j(c^{\ell'}_{j}) \leq C_j(c^1_{i}) + C_j(c^2_{i})\). Combining the inequalities we have $\sum_{t = 2}^{\ell'} C_j(c^t_j) \le \sum_{t = 1}^{\ell'} C_j(c^t_i) \le C_j(c^1_j) + \sum_{t = 1}^{\ell'} C_j(c^t_i)$ and agent $j$ does not strongly envy agent $i$. Therefore, no agent strongly envies any other agent, and we have a tEFX allocation. Clearly, the round-robin procedure takes polynomial time after sorting the chores beforehand, which also takes polynomial time. Therefore, we have a polynomial time algorithm that outputs a tEFX allocation.
\end{proof}

Applying Lemma~\ref{tEFX for three agents lemma} for $\ell = \lceil \frac{m}{n} \rceil$ gives a full allocation $X$ that is tEFX for agents with additive 2-ratio-bounded cost functions.

The round-robin algorithm runs in polynomial time since we only need to sort the items based on the cost function of each agent. Thus from Lemmas~\ref{(1 + (alpha - 1)/(l-1)-EFX chores} and \ref{tEFX for three agents lemma} we conclude the proof of Theorems~\ref{alpha ratio theorem} and \ref{tEFX 2-ratio}. The following statement is derived as a corollary of Theorem~\ref{alpha ratio theorem}.


\begin{corollary}
    There exists a 2-EFX allocation for chores among any number of agents with additive $\alpha$-ratio-bounded cost function when $m \ge \alpha n$. Additionally, it can be computed in polynomial time.
\end{corollary}
\section{Conclusion and Future Work}
In this paper, we gave polynomial time algorithms for the computation of 2-EFX allocations of chores for three agents with subadditive cost functions. However, our proof relies on an extensive case study; we raise the question of whether a more elegant proof exists. Moreover, we showed that tEFX allocations exist for $n$ agents split into three groups based on their cost functions, one group with identical general cost functions, one group with identical and additive 2-ratio bounded cost functions, and one group consisting of one agent with a general cost function. For $n$ agents and $m$ chores, we proved the existence of tEFX allocations for agents with additive 2-ratio-bounded cost functions and $(1 + \frac{\alpha - 1}{\lceil \frac{m}{n} \rceil  -1})$-EFX for $n$ agents with additive $\alpha$-ratio-bounded cost functions. Future research could aim to enhance approximation guarantees for three agents with additive, subadditive, or even more general cost functions or confirm the existence of tEFX allocations under similar settings.
Additionally, exploring $\alpha$-EFX allocations of chores involving $n$ agents, where $\alpha$ is  a constant, could extend our current methodologies. An alternative avenue of exploration could involve improving approximation guarantees for $n$ agents with additive and $\alpha$-ratio-bounded cost functions. This could include either achieving a 2-EFX allocation when the total number of chores is fewer than $\alpha n$ or establishing the existence of an EFX allocation for $n$ agents who have additive and 2-ratio-bounded cost functions.

\bibliographystyle{apalike}
\bibliography{main}
\begin{appendix}
\section{Sufficiency of Non-degenerate Instances}
\label{appen_non_degenerate}

We show it is sufficient to establish the existence of EFX allocations solely for non-degenerate instances. A cost function $C_i$ is \emph{non-degenerate} if $C_i(S) \neq C_i(T)$ whenever $S \neq T$. It is well known that cost functions can be assumed to be non-degenerate in the goods' case, see for example~\cite{DBLP:journals/corr/abs-2205-07638} and~\cite{chaudhury2020efx}. An identical argument also works for chores. We first note that it has been shown in~\cite{ijcai2023p0277} that it is sufficient to consider non-degenerate instances when considering tEFX allocations for chores. We now give similar arguments for EFX and $\alpha$-EFX allocations. 


Let $$\delta = \min{}_{S, T \subseteq [m]: C_i(S) \neq C_i(T)} |C_i(S) - C_i(T)|,$$ and let $\epsilon > 0$ be such that $\epsilon \cdot 2^{m+1} < \delta$. Define $C'_i(S) = C_i(S) + \epsilon \sum_{j \in S} 2^j$ for every $S \subseteq [m]$.

\renewcommand\thetheorem{15}
\begin{lemma}\label{inequality for non-degenerate lemma}
    For $S, T \subseteq [m]$ such that $C_i(S) > C_i(T)$, we have $C'_i(S) > C'_i(T)$.
\end{lemma}
\begin{proof}
We have 
    \begin{align*}
    \abs{C'_i(S) - C'_i(T)} & = \abs{C_i(S) - C_i(T) + \epsilon (\sum_{j \in S} 2^j - \sum_{j \in T} 2^j)} \\
                      & \geq \delta - \epsilon (2^{m+1} - 1) \\
                      &> 0.
\end{align*}
\end{proof}

\begin{lemma}\label{non-degenerate sufficiency for exf}
$C'_i$ is non-degenerate. Furthermore, if $X = (X_1, X_2, \dots, X_n)$ is an EFX allocation for $ \mathcal{C'} = (C'_1,\ldots,C'_n)$ then $X$ is also an EFX allocation for $\mathcal{C}$.
\end{lemma}
\begin{proof}
    Consider any two sets $S, T \subseteq [m]$ such that $S \neq T$. If $C_i(S) \neq C_i(T)$, we have $C'_i(S) \neq C'_i(T)$ by Lemma~\ref{inequality for non-degenerate lemma}. If $C_i(S) = C_i(T)$, we have $$C'_i(T) - C'_j(S) = \epsilon(\sum_{j \in T} 2^j - \sum_{j \in S} 2^j) \neq 0.$$ So, $C'_i$ is non-degenerate.

    For the final claim, assume that $X$ is an EFX allocation for $\mathcal{ C'}$. Let $i, j \in [n]$ and $c \in X_i$. Then $C'_i(X_i \setminus c) < C'_i(X_j)$; since $C'_i$ is non-degenerate, we cannot have equality. Thus  $C_i(X_i \setminus c) \le C_i(X_j)$ by Lemma~\ref{inequality for non-degenerate lemma}.
\end{proof}

A similar argument to Lemma~\ref{non-degenerate sufficiency for exf} goes for $\alpha$-EFX allocations.
\renewcommand\thetheorem{17}
\begin{lemma}\label{non-degenerate sufficiency for alpha-exf}
If $X = (X_1, X_2, \dots, X_n)$ is an $\alpha$-EFX allocation for $\mathcal{ C'} = (C'_1,\ldots,C'_n)$ then $X$ is also an $\alpha$-EFX allocation for $\mathcal{ C}$.
\end{lemma}
\begin{proof}
    Assume that $X$ is an $\alpha$-EFX allocation for $\mathcal{ C'}$. Let $i, j \in [n]$ and $c \in X_i$. Then $C'_i(X_i \setminus c) < \alpha \cdot C'_i(X_j)$; since $C'_i$ is non-degenerate, we cannot have equality. Thus,  $C_i(X_i \setminus c) \le \alpha \cdot C_i(X_j)$ by Lemma~\ref{inequality for non-degenerate lemma}.
\end{proof}

\end{appendix}

\end{document}